\documentclass[11pt,a4paper]{article}

\usepackage[a4paper, left=1in, right=1in, top=1in, bottom=1in, includehead, includefoot, foot = 0pt, head=0pt]{geometry} 
\usepackage[group-separator={,},group-digits=integer]{siunitx} 
\usepackage{booktabs} 
\usepackage{enumitem} 

\usepackage[round,colon,authoryear]{natbib} 
\usepackage[pdftex,colorlinks,citecolor=blue,urlcolor=blue,linkcolor=red]{hyperref} 
\usepackage{graphicx} 
\parskip=\medskipamount 
\usepackage[small,bf,hang]{caption}	
\usepackage{amssymb} 
\usepackage{amsmath} 
\usepackage{bbm} 
\usepackage{amsbsy} 
\usepackage{mathrsfs} 
\usepackage{multirow} 
\usepackage[dvipsnames]{xcolor} 
\usepackage{subcaption} 
\usepackage[hang,flushmargin]{footmisc} 
\usepackage{tikz} 
\usetikzlibrary{arrows,decorations.pathmorphing,backgrounds,positioning,fit,petri,patterns}
\usetikzlibrary{decorations.pathreplacing}
\usetikzlibrary{matrix} 
\usepackage{rotfloat}
\usepackage{titling}
\usepackage{mathtools}

\usepackage[ruled]{algorithm2e} 
\SetAlCapFnt{\small} 
\SetKw{KwTo}{in}

\usepackage[official]{eurosym}
\usepackage{soul}

\DeclareMathOperator{\logit}{logit}

\usepackage{authblk}

\usepackage{fancyhdr}
\graphicspath{{"Images/"},{"../R code/Images/"}, {"../../Research RBNS/R code/Images/"}}

\newcommand{\vecx}{\boldsymbol{x}}

\newcommand{\bs}[1]{\boldsymbol{#1}}

\newcommand{\tr}[1]{}

\usepackage{bbm}

\usepackage{environ}

\begin{document}
\title{Bridging the gap between pricing and reserving with an occurrence and development model for non-life insurance claims}

\author[1]{Jonas Crevecoeur}
\author[2,3,4,5]{Katrien Antonio}
\author[6]{Stijn Desmedt}
\author[6]{Alexandre Masquelein}
\affil[1]{Interuniversity Institute for Biostatistics and statistical Bioinformatics (I-BioStat), Data Science Institute, Hasselt University, Belgium.}
\affil[2]{Faculty of Economics and Business, KU Leuven, Belgium.}
\affil[3]{Faculty of Economics and Business, University of Amsterdam, The Netherlands.}
\affil[4]{LRisk, Leuven Research Center on Insurance and Financial Risk Analysis, KU Leuven, Belgium.}
\affil[5]{LStat, Leuven Statistics Research Center, KU Leuven, Belgium.}
\affil[6]{QBE Re, Belgium}
\date{\today}
\maketitle
\thispagestyle{empty}

\begin{abstract}
Due to the presence of reporting and settlement delay, claim data sets collected by non-life insurance companies are typically incomplete, facing right censored claim count and claim severity observations. Current practice in non-life insurance pricing tackles these right censored data via a two-step procedure. First, best estimates are computed for the number of claims that occurred in past exposure periods and the ultimate claim severities, using the incomplete, historical claim data. Second, pricing actuaries build predictive models to estimate technical, pure premiums for new contracts by treating these best estimates as actual observed outcomes, hereby neglecting their inherent uncertainty. We propose an alternative approach that brings valuable insights for both non-life pricing as well as reserving. As such we effectively bridge these two key actuarial tasks that have traditionally been discussed in silos. Hereto we develop a granular occurrence and development model for non-life claims that tackles reserving and at the same time resolves the inconsistency in traditional pricing techniques between actual observations and imputed best estimates. We illustrate our proposed model on an insurance as well as a reinsurance portfolio.The advantages of our proposed strategy are most compelling in the reinsurance illustration where large uncertainties in the best estimates originate from long reporting and settlement delays, low claim frequencies and heavy (even extreme) claim sizes.

\end{abstract}

\paragraph{Keywords:} non-life pricing; non-life reserving; statistical and machine learning methods; reinsurance; occurrence, reporting and development of claims

\section{Introduction} \label{section:claim_development_model}
The insurance industry is characterized by an inverted production cycle in which the premium for a new contract has to be determined before observing the associated loss. Pricing actuaries estimate the technical price of a cover by modelling historical loss data. In non-life insurance, the total loss $L$ on a new contract is often estimated via a frequency-severity decomposition \citep{Denuit2007, Frees2008}, which models the expected loss as 
$$E(L) = E(N) \cdot E(Y),$$
assuming independence between the number of occurred claims $N$ and their severity $Y$.
Risk-based premiums then follow by taking risk characteristics into account when building predictive models for the historical claim frequency and severity data. Pricing requires a data set with claim counts registered at the level of individual contracts and ultimate claim sizes at the level of individual claims. 

\begin{figure}
\center
\begin{tikzpicture}[scale=0.7, auto, to/.style={->,>=stealth',shorten >=1pt}, every node/.style={font=\fontsize{9pt}{9pt}\selectfont\sffamily, align=center, semithick}]

\draw[to] (-.5,0) -- (13,0) node[below right]  {Time};

\draw[to] (1,3) -- (1,1);
\draw (1,3.5) node[baseline] {Occurrence};
\draw[to] (4,3) -- (4,1);
\draw (4,3.5) node[baseline] {Reporting};

\draw[to] (11,3) -- (11,1);
\draw (11,3.5) node {Settlement};

\node[circle,draw, inner sep=.3cm,
           path picture={
               \node at (path picture bounding box.center){
                   \hspace{0.12cm}\includegraphics[width=.8cm]{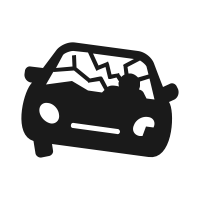}
               };
           },
			fill=white] at (1, 0) {};
			
\node[circle,draw, inner sep=.3cm,
           path picture={
               \node at (path picture bounding box.center){
                   \hspace{0.12cm}\includegraphics[width=.5cm]{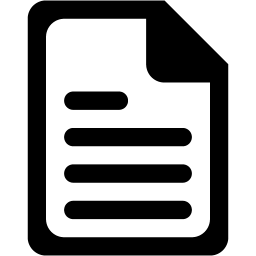}
               };
           },
			fill=white] at (4, 0) {};

\node[circle,draw, inner sep=.3cm,
           path picture={
               \node at (path picture bounding box.center){
                   \hspace{0.12cm}\includegraphics[width=.5cm]{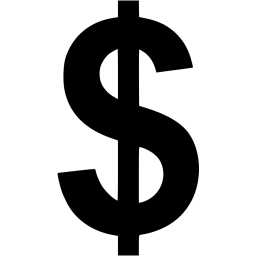}
               };
           },
			fill=white] at (6, 0) {};

\node[circle,draw, inner sep=.3cm,
           path picture={
               \node at (path picture bounding box.center){
                   \hspace{0.12cm}\includegraphics[width=.5cm]{iconPayment.png}
               };
           },
			fill=white] at (7.5, 0) {};	

\node[circle,draw, inner sep=.3cm,
           path picture={
               \node at (path picture bounding box.center){
                   \hspace{0.12cm}\includegraphics[width=.5cm]{iconPayment.png}
               };
           },
			fill=white] at (9, 0) {};			

\node[circle,draw, inner sep=.3cm,
           path picture={
               \node at (path picture bounding box.center){
                   \hspace{0.12cm}\includegraphics[width=.8cm]{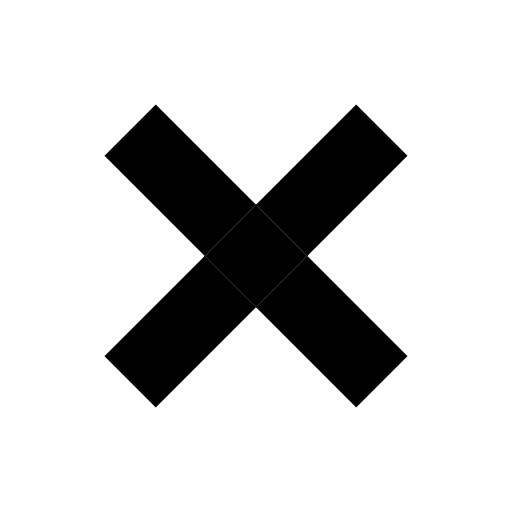}
               };
           },
			fill=white] at (11, 0) {};


tikz={\draw [decorate, line width=1pt, decoration={brace, mirror, amplitude=3mm}] (4, -0.8) -- (11, -0.8) node [midway, below=4mm] {Settlement delay} ;}

tikz={\draw [decorate, line width=1pt, decoration={brace, mirror, amplitude=3mm}] (1, -0.8) -- (4, -0.8) node [midway, below=4mm] {Reporting delay} ;}

\node[] at (0, -2.5) {};

\end{tikzpicture}
\caption{Development process of a single claim}
\label{figure:development_complete}
\end{figure}

Figure~\ref{figure:development_complete} visualizes the development process of a single claim. This process starts with the occurrence of an insured event, which is reported to the insurer after some delay. If the claim is eligible for compensation under the insurance policy, a number of payments follow. Finally, the claim settles and we observe its total cost. Depending on the insurer and line of business other relevant events (e.g.,~the involvement of a lawyer) will be registered during the lifetime of a claim. For claims that settled before the moment of evaluation, the reserving actuary observes the full development process and thus the total claim size. However, the development process is only partially observed for reported, but not yet settled claims. For claims that occurred in the past, but are not yet reported the entire development process is missing in the insurer's database. Such reporting and settlement delays are particularly relevant in long-tailed business lines (e.g.,~workers' compensation insurance or reinsurance contracts) where claim settlement can take several years.

Due to the delays present in the claim development process, the pricing actuary only observes the number of reported claims instead of the total number of claims that occurred in past exposure periods. Similarly, the amounts already paid for open claims underestimate actual, ultimate losses, since future payments are missing. As a result of the incomplete claim history, pricing requires a two step approach. First, claim counts and sizes are estimated per policy and per claim, respectively, based on the available claim history, i.e.
\begin{align*}
	\hat{N} = E(N \mid \mathcal{F}_\tau) \quad \text{and} \quad \hat{Y} = E(Y \mid \mathcal{F}_\tau),
\end{align*}
where $\mathcal{F}_{\tau}$ denotes the information available at the evaluation or observation date $\tau$.  In a second step, these estimates, so called best-estimates, are treated as actual observations when the pricing actuary constructs predictive models for claim frequency and severity as a function of risk characteristics.

In practice, pricing actuaries may ignore the first step of this pricing procedure and only consider reported and settled claims. This approach is feasible when reporting and settlement delays are small and limited bias is introduced by ignoring the censoring present in the data. Alternatively, the best estimates in the first step of the pricing procedure can be constructed in several ways. Claim handlers may estimate the number of unreported claims per policy and the future claim costs based on their expert opinion. Combined with the amount already paid, the estimate of the future cost on a claim then constitutes the expert's best estimate of the ultimate claim size, also called the incurred claim amount. As a data-driven alternative, methods from non-life reserving can be adapted to estimate the total, ultimate cost of individual claims as well as the number of occurred, but not reported claims. The literature on non-life reserving unravels along two axes: aggregate and individual reserving models. Aggregate reserving models (e.g.,~the chain ladder method \citep{Mack1993, Mack1999}) ignore individual claim characteristics and model a single claim development process for all claims that occur within an accident year. Best estimates for pricing are then obtained by applying this (aggregate) development pattern to individual, reported claims. Constructing these best estimates from an aggregate reserving model has two important disadvantages. First, most aggregate reserving models do not distinguish between open and settled claims. Consequently, the development pattern is estimated from a mix of both open and closed claims, and is then applied to both types of claims. While the best estimate of a settled claim will differ from its true, observed cost, it should replace the true, observed value to be consistent with the aggregate reserving model. Second, ignoring risk characteristics when constructing best estimates diminishes the heterogeneity that is present in the claim severity data analyzed by pricing actuaries. Following \cite{Norberg1993, Norberg1999} a literature on individual reserving models has emerged, where best estimates are constructed at the level of individual claims. We see most potential in a stream of individual reserving models in discrete time adopting techniques from the insurance pricing literature. \cite{Larsen2007}, \cite{Wuthrich2018}, \cite{Crevecoeur2020} and \cite{Delong2020} focus on generalized linear models (GLMs), regression trees, gradient boosting models and neural networks for claims reserving, respectively. In these approaches the inclusion of claim-specific covariates tailors the best estimates to the characteristics of the individual claims. Consequently, the completed data sets will more accurately reflect the heterogeneity in the claim data. To the best of our knowledge no data driven methods have been published for estimating the number of unreported claims at the level of individual policies.

Insurance pricing literature mainly puts focus on the second step of the pricing procedure, where a statistical model is fitted to the best estimates. Although actual observations and best estimates follow different statistical distributions, the frequency-severity decomposition still holds, i.e.
\begin{align*}
	E(L) = E(N) \cdot E(Y) = E\big(E(N \mid \mathcal{F}_\tau)\big)  \cdot E\big(E(Y \mid \mathcal{F}_\tau)\big),
\end{align*}
as a result of the tower rule. This property is essential, since it enables an unbiased estimate of the loss from predictive models calibrated on the best estimate claims data. However, many other properties of the loss (e.g.,~the variance) are not preserved when treating best estimates as actual observations. In particular, severity is underestimated for policies covering losses above a (known) deductible $D$. This is a consequence of Jensen's inequality \citep{jensen1906}, which states that $ E[\varphi(Y) \mid \mathcal{F}_{\tau}] \geq \varphi(E(Y \mid \mathcal{F}_{\tau})) $ for any random variable $Y$ and convex function $\varphi(.)$. Indeed, applied to an insurance contract with deductible $D$, we obtain
$$E((Y-D)_+) = E[E((Y-D)_+ | \mathcal{F}_\tau)] \geq E\left[ (E(Y \mid \mathcal{F}_\tau)- D)_+\right],$$
for the convex function $\varphi: Y \to (Y - D)_+$. This is especially relevant in excess-of-loss reinsurance pricing, where deductibles are high and long settlement delays result in many open claims. Moreover, the risk characteristics selected when modelling frequency and severity data and their calibrated effects should rather be interpreted as effects on best estimates instead of effects on actual observations. These effects are likely to be affected by the method used for constructing these best estimates.

Our paper contributes by proposing a novel approach for non-life insurance pricing that resolves the inconsistencies between actual observations and best estimates in traditional pricing. Moreover, by modelling the occurrence and development process of claims, our proposed model is also readily available for reserving. Hence, we bridge two key tasks of the non-life actuary that are typically studied in silos. We demonstrate our methodology with a case-study on pricing and reserving for both a traditional insurance as well as a reinsurance portfolio. This is one of the first papers applying techniques from individual reserving on a reinsurance data set. The reinsurance industry is characterized by low claim frequencies and large claim severities \citep{Albrecher2017}, which demands special attention when building predictive models for the development of individual claims.

This paper is organized as follows. Section~\ref{section:odm} introduces our proposed model for the occurrence and development of claims at the level of an individual non-life insurance contract. Section~\ref{section:apply_odm} illustrates how this model can be used for pricing and reserving with non-life insurance policies. Section~\ref{section:casestudy} demonstrates this methodology on two case studies. Section~\ref{section:conclusion} concludes the paper.

\section{An occurrence and development model for non-life insurance claims} \label{section:odm}

We present a discrete time occurrence and development model (ODM). This ODM captures the occurrence and the reporting of claims at the level of an individual insurance contract, as explained in Section~\ref{section:reporting}. Section~\ref{section:hierarchical} details how the ODM structures the development of individual claims after reporting. Together these two building blocks drive the complete development of all claims occurring on a portfolio of insurance contracts. In the remainder of this paper, we implicitly assume a yearly, discrete time grid. However, our approach extends directly to quarterly, monthly or daily time grids.

\subsection{Modelling the occurrence and reporting of non-life claims} \label{section:reporting}

We consider a portfolio with historical claims data registered on $n$ policies. Each of these policies covers the claims occurring during a single year of exposure.\footnote{In case the historical data set contains policies covering multiple occurrence years, these policies are split into multiple policies that each cover a single occurrence year to match the required data format.} Let $N_i$ denote the claim frequency on policy $i$, i.e.~the total number of claims that occur in the occurrence year, $\texttt{occ(i)}$, covered by this policy. Due to a possible delay in reporting (see Figure~\ref{figure:development_complete}), these counts $N_i$ are not directly observable. Instead we observe counts $N_{ij}$, which register the number of claims from policy $i$ that are reported in the $(j-1)$-th year after occurrence, i.e.~in year $\texttt{occ(i)} + j - 1$. At the observation date $\tau$, the set of observed claims consists of $\{N_{ij} \mid i = 1, \ldots, n, j = 1, \ldots, \tau_i \}$ where $\tau_i \coloneqq \max(d, \tau - \texttt{occ}(i) + 1)$ is the number of observed reporting years for policy $i$ with $d$ denoting the maximal reporting delay. The set of not (yet) reported claims consists of $\{N_{ij} \mid i = 1, \ldots, n, j = \tau_i + 1, \ldots, d \}$.  Following \cite{Jewell1990} and \cite{Norberg1993}, we propose a model to predict these unreported claim counts, based on the following assumptions:
\begin{itemize}
 \item[(F1)] Claims are reported with a maximal delay of $d$ years. This maximal delay $d$ is at most the length of the observation window $\tau$ of the portfolio, i.e. $d \leq \tau$.
 \item[(F2)] Conditional on the observed policy covariates $\boldsymbol{x}_i$, claim counts $N_i$ (with $i=1,\ldots,n$) are independent and follow a Poisson distribution with intensity $\lambda(\boldsymbol{x}_i)$.
 \item[(F3)] Conditional on the total number of claims $N_i$ on policy $i$ and its covariates $\boldsymbol{x}_i$, the reported claim counts $N_{ij}$ are multinomially distributed with reporting probabilities $p_{j}(\boldsymbol{x}_i)$, where $j=1,\ldots,d$.
\end{itemize}
Assumption (F1) limits the reporting delay and allows to retrieve the total claim frequency on policy $i$ as
$$N_i = \sum_{j=1}^d N_{ij}.$$
The independence assumptions in (F2-F3) are similar to those in classical insurance pricing, but might be violated in case of high impact events, e.g.~extreme weather with claims occurring in clusters, requiring more advanced modelling techniques to capture dependencies. As a result of the thinning property for Poisson distributions, assumptions (F2-F3) imply
$$
	N_{ij} \sim \text{POI}(\lambda(\boldsymbol{x}_i) \cdot p_{j}(\boldsymbol{x}_i)).
$$
The log-likelihood of the observed claim counts then follows as
\begin{equation} \label{eq:likelihood_occ_rep_full}
 \mathcal{L}(\bs{\lambda}, \bs{p}) =\sum_{i=1}^{n} \sum_{j=1}^{\tau_i} - \lambda(\boldsymbol{x}_i) \cdot p_{j}(\boldsymbol{x}_i) + N_{ij} \cdot \log(\lambda(\boldsymbol{x}_i)) + N_{ij} \cdot \log( p_{j}(\boldsymbol{x}_i)) - \log(N_{ij}!),
\end{equation}
where $\boldsymbol{\lambda}$ and $\boldsymbol{p}$ are a shorthand notation for the parameters used in the Poisson intensities and reporting probabilities.
Extending the work of \cite{Verbelen2017} designed for claims reserving, we now specify the above likelihood at the level of individual policies, with a tailored specification for the reporting of claims via the reporting probabilities in $\bs{p}$. The joint estimation of $\bs{\lambda}$ and $\bs{p}$ in $\eqref{eq:likelihood_occ_rep_full}$ is complicated by the presence of the interaction term $\lambda(\boldsymbol{x}_i) \cdot p_{j}(\boldsymbol{x}_i)$. Using an EM-algorithm \citep{Dempster1977}, the occurrence and reporting parameters in $\eqref{eq:likelihood_occ_rep_full}$ can be decoupled and estimated iteratively. The $k$-th expectation (E) step then imputes the hidden observations $\{N_{ij} \mid i \leq n,  \tau_i < j \leq d\}$ as follows:
$$
N_{ij}^{(k)} = \begin{cases}
	N_{ij} & j \leq \tau_i \\
	\lambda^{(k-1)}(\boldsymbol{x}_i) \cdot p_{j}^{(k-1)}(\boldsymbol{x}_i) & \tau_i < j \leq d,
\end{cases}
$$
where the superscript $(k-1)$ refers to the parameter estimates obtained in the previous iteration of the EM-algorithm. The $k$-th maximization (M) step then maximizes the completed log-likelihood
\begin{eqnarray*}
&&\mathcal{L}_c(\bs{\lambda}^{(k)}, \bs{p}^{(k)}) = \\ && \sum_{i=1}^{n} \left[ -\lambda^{(k)}(\boldsymbol{x}_i) + N_{i}^{(k-1)} \cdot \log(\lambda^{(k)}(\boldsymbol{x}_i)) + \sum_{j=1}^d \left\{ N^{(k-1)}_{i,j} \cdot \log( p_{j}^{(k)}(\boldsymbol{x}_i)) - \log(N^{(k-1)}_{ij}!)\right\}\right],
\end{eqnarray*}
where $N^{(k-1)}_i = \sum_{j=1}^d N^{(k-1)}_{ij}$. The likelihood now splits in an occurrence and reporting contribution. For the occurrence process, we maximize
$$ \mathcal{L}_c^{\texttt{occ}}(\bs{\lambda}^{(k)}) = \sum_{i=1}^{n} \left[- \lambda^{(k)}(\boldsymbol{x}_i) + N^{(k-1)}_{i} \cdot \log(\lambda^{(k)}(\boldsymbol{x}_i))\right].$$
This likelihood is proportional to the Poisson likelihood that is typically optimized in the claim frequency models used in insurance pricing. The partially observed claim counts $N_i$ are replaced by counts $N^{(k-1)}_i$, adjusted for unreported claims. For the reporting process, we maximize 
\begin{equation} \label{eq:likelihood_reporting}
 \mathcal{L}_c^{\texttt{rep}}(\bs{p}^{(k)}) =  \sum_{i=1}^{n} \sum_{j=1}^d N^{(k-1)}_{ij} \cdot \log( p^{(k)}_{j}(\boldsymbol{x}_i)), \quad \text{subject to} \quad \sum_{j=1}^d  p^{(k)}_{j}(\boldsymbol{x}_i) = 1, \forall i.
\end{equation}
The estimation of the reporting probabilities $p_{ij}:=p^{(k)}_{j}(\boldsymbol{x}_i)$ in this multinomial likelihood is complicated by the sum-to-one restriction on the reporting probabilities that must hold for each policy $i$. Following \cite{KalbfleischLawless1991}, we overcome the sum-to-one restriction by projecting the $d$ probabilities $(p_{ij})_{j=1\ldots,d}$ into $d-1$ probabilities $(q_{ij})_{j=1\ldots,d-1}$ as follows
\begin{align*}
	q_{ij} &= P(\texttt{rep.delay} = j + 1 \mid \texttt{rep.delay} \leq j + 1) \\
			&= \frac{p_{i, j + 1}}{\sum_{\kappa = 1}^{j+1} p_{i, \kappa}} \quad \text{for}\ j = 1, \ldots, d-1.
\end{align*}
The $q$ probabilities take the form of inverted, discrete time hazard rates from which the vector of probabilities $(p_{ij})_{j=1\ldots,d}$ can be retrieved as
\begin{equation}
p_{ij} = \begin{cases}
	\prod_{\kappa = 1}^{d-1} (1-q_{i, \kappa}) & j = 1 \\
	q_{i, j-1} \cdot \prod_{\kappa = j}^{d-1} (1-q_{i, \kappa}) & 1 < j < d\\
	q_{i, d-1} & j = d
\end{cases}. \label{eq:reparametrize_q}
\end{equation}

Combining \eqref{eq:reparametrize_q} with \eqref{eq:likelihood_reporting} and changing the order of summation, the likelihood for the reporting process becomes
\begin{equation} \label{eq:likelihood_reporting_q}
 \mathcal{L}_c^{\texttt{rep}}(\bs{q}^{(k)}) =  \sum_{i = 1}^n \sum_{\kappa=1}^{d-1} \left( \sum_{j = 1}^{\kappa} N^{(k-1)}_{ij}\right) \cdot \log(1-q_{\kappa}^{(k)}(\boldsymbol{x}_i)) + \sum_{i = 1}^n \sum_{j=1}^{d-1} N^{(k-1)}_{i, j+1} \cdot \log(q^{(k)}_{j}(\boldsymbol{x}_i)).
\end{equation}
This likelihood is a sum of binomial likelihood contributions and can be optimized with standard statistical modelling techniques.

When applied for pricing, the proposed occurrence and reporting model estimates the expected number of claims $\hat{\lambda}(\boldsymbol{x}_i)$ per policy while correcting for the existence of unreported claims that occurred in the exposure year covered by the policy. When used for reserving, the model estimates the number of claims that will be reported in future years on policy $i$, i.e., $(N_{i, \tau_i+1}, \dots, N_{i, d})$, as well as their associated reporting delays. Estimating unreported claims at policy level has the advantage that policy specific reserves can be booked for these claims.

\subsection{A hierarchical model for the development of reported non-life claims} \label{section:hierarchical}

Insurers track many dynamic claim characteristics (e.g.,~the amount paid, settlement status, involvement of a lawyer) over the lifetime of a claim. We predict the joint evolution of these dynamic claim characteristics using a hierarchical individual claims reserving model originally  proposed in \citep{Crevecoeur2020}. This section restates the key features of this model and proposes some extensions. For a more in depth analysis of the hierarchical reserving model we refer to our original paper. 

When modelling the development of a claim, we will differentiate between the initial state of the claim characteristics as observed in the reporting year of the claim and the updates in later years. This distinction reflects the difference in information that becomes available in the reporting year compared to later years. For example, in the reporting year the claim expert sets the initial incurred amount, which can be quite large, while in later development years small adjustments to this initial incurred are observed. We let the vector $\boldsymbol{I}_k$ structure the initial claim characteristics for claim $k$ as registered at the end of its reporting year, denoted $\texttt{rep(k)}$. In later years, update vectors $\boldsymbol{U}_k^j$ (with $j \geq 2$) structure the evolution of claim $k$ in the $(j-1)$-th year since reporting, i.e.~in year $\texttt{rep(k)} + j - 1$. The information captured by the vectors $\boldsymbol{I}_k$ and $\boldsymbol{U}_k^j$ is tailored to the portfolio at hand. The case-studies in Section~\ref{section:casestudy} illustrate a possible set up in which the joint evolution of the settlement status, the amount paid and the incurred are tracked over the lifetime of a claim. We refer to these chosen characteristics as the layers of the hierarchical reserving model. Let the vector $\mathcal{X}_k$ store the observed development of claim $k$, i.e.~
$$ \mathcal{X}_k \coloneqq \{ \boldsymbol{I}_k, \boldsymbol{U}_k^2, \ldots, \boldsymbol{U}_k^{\tau_k}\},$$
with $\tau_k = \tau - \texttt{rep(k)} + 1$ the number of observed years since reporting for claim $k$. Our approach models the development of claim $k$ as recorded in $\mathcal{X}_k$ based on a single assumption: 
\begin{itemize}
	\item[(S1)] Conditional on static claim covariates $\vecx_k$ available at the reporting of claim $k$, the development of the claim is independent of the development of the other claims in the portfolio.
\end{itemize}
This independence assumption is essential for modelling the development at the level of individual claims. As a result of (S1) we can write the likelihood for a portfolio with $m$ reported claims as
$$ \mathcal{L} = \prod_{k = 1}^m f(\boldsymbol{I}_k, \boldsymbol{U}_k^2, \ldots, \boldsymbol{U}_k^{\tau_k} \mid \vecx_k),$$
where $f(\boldsymbol{I}_k, \boldsymbol{U}_k^2, \ldots, \boldsymbol{U}_k^{\tau_k} \mid \vecx_k)$ is the joint likelihood of the development process observed for claim $k$. Our hierarchical approach decomposes this joint likelihood over time as well as over the layers (i.e.~the respective dimensions) of the vectors $\boldsymbol{I}_k$ and $\boldsymbol{U}_k^j$ by applying the law of conditional probability twice.
First, the likelihood is split in chronological order
\begin{align*}
	\mathcal{L} = \prod_{k = 1}^m f(\boldsymbol{I}_k \mid \vecx_k) \cdot \prod_{j=2}^{\tau_k} f(\boldsymbol{U}_k^j \mid \boldsymbol{I}_k, \boldsymbol{U}_k^2, \ldots, \boldsymbol{U}_k^{j-1},  \vecx_k).
\end{align*}
By conditioning on past events, we allow the model to use the historical development of a claim (e.g.,~total amount paid, reserve, settlement status in previous years) when modelling the development in future years. Second, we decompose the likelihood over the layers of $\boldsymbol{I}_k$ and $\boldsymbol{U}_k^j$
\begin{align*}
\mathcal{L} & = \prod_{k = 1}^m \prod_{l=1}^v f(I_{k,l} \mid I_{k,1}, \ldots, I_{k, l-1},  \vecx_k) \\
						 & \phantom{={}} \cdot \quad  \prod_{k = 1}^m \prod_{j=2}^{\tau_k} \prod_{l=1}^w f(U_{k,l}^j \mid \boldsymbol{I}_k, \boldsymbol{U}_k^2, \ldots, \boldsymbol{U}_k^{j-1}, U_{k,1}^j, \ldots, U_{k,l-1}^j,  \vecx_k),
\end{align*}
where $v$ and $w$ denote the length (i.e., the number of layers) of the initial vector $\boldsymbol{I}_k$ and update vector $\boldsymbol{U}_k^j$ respectively. Through conditioning on the layered structure, we allow for dependencies in the development of the claim characteristics within a time period. We model this decomposed likelihood by specifying a statistical model per layer, leading to a total of $v+w$ statistical models.

When applied to pricing, we use the proposed hierarchical development model to estimate the total severity of claims. When used for reserving purposes, the model allows to estimate the future cost of reported as well as not yet reported claims, while accounting for their static characteristics registered at reporting as well as their observed development so far.

\section{Pricing and reserving with the occurrence and development model} \label{section:apply_odm}

\subsection{Non-life pricing} \label{section:pricing}
Following the frequency-severity decomposition discussed in Section~\ref{section:claim_development_model}, we estimate the pure premium $\pi_i$ for policy $i$ as the product of its expected claim frequency, $E(N_i)$, and expected claim severity, $E(Y_i)$, i.e.~
$$\pi_i = E(N_i) \cdot E(Y_i).$$
Risk-based claim frequency estimates follow immediately from the occurrence and reporting model proposed in Section~\ref{section:reporting}. In contrast with traditional claim frequency models, our approach adjusts the estimated claim frequencies for the presence of unreported claims. We consider two strategies for modelling the distribution of the claim severity given a set of policy covariates $\boldsymbol{x}_i$ with our ODM. The first approach simulates new claims for a given policy from ground up, whereas the second approach simulates the future development of already reported open claims.

\paragraph{Simulating new claims} We use the ODM calibrated on historical claims data to simulate the ultimate cost of a large number of new claims occurring on a given policy.  Algorithm~\ref{algorithm:full_simulation} outlines the procedure to simulate the occurrence, reporting and development of a new claim on a policy with characteristics $\boldsymbol{x}$.

\begin{algorithm}[H]
	\DontPrintSemicolon
	\SetKw{KwSimulate}{Simulate}
	\SetKw{KwSet}{Set}
	\SetKw{KwEvaluate}{Evaluate}
	\SetKwRepeat{Do}{do}{while}
    \SetAlgoLined
    \KwIn{policy with characteristics $\boldsymbol{x}$}
    \KwOut{simulation of the ultimate claim severity}
    \;
    	\KwSimulate $\texttt{rep.delay}$ from the reporting distribution $\boldsymbol{p}(\boldsymbol{x})$. \\
    	\KwSimulate $\boldsymbol{I}$ given $\boldsymbol{x}, \texttt{rep.delay}$. \\
    	\KwSet $s = 1$. \\
    	\If{not.settled($\boldsymbol{I}$)} {
    		\Do{not.settled($\boldsymbol{I},\boldsymbol{U}^2, \ldots, \boldsymbol{U}^s$)}{
    			\KwSimulate $\boldsymbol{U}^{s+1}$ given $\boldsymbol{x}, \texttt{rep.delay},\boldsymbol{I},\boldsymbol{U}^2, \ldots, \boldsymbol{U}^s$. \\
    			\KwSet $s = s + 1$. \\
    		}
    	}
		\KwEvaluate $Y = Y(\boldsymbol{I}, \boldsymbol{U}^2, \ldots, \boldsymbol{U}^s)$.
    \caption{Simulating the ultimate severity of a new claim}
    \label{algorithm:full_simulation}
\end{algorithm}

It is essential that the paid amount or the incurred is tracked within $\boldsymbol{I}$ and $\boldsymbol{U}^{j}$, such that the claim's ultimate cost at settlement can be computed as a function of the simulated development process. Using these simulated paths we obtain an empirical distribution of a claim's ultimate cost from which the expected severity follows. 

\paragraph{Simulating future paths for open claims} In this alternative modelling strategy we first simulate for each open claim a large number of future paths, say $n_{\texttt{path}}$. Each simulated path $p$ of an open claim $k$ corresponds to a scenario for the ultimate claim size $Y_{k, p}$. Combining these simulated paths we obtain a distribution of the ultimate size per claim. In a second step, we fit a severity distribution by assigning a weight of one to actual observations from closed claims and a weight of $\frac{1}{n_{\texttt{path}}}$ to the ultimate claim sizes corresponding to the simulated paths for the open claims, i.e. we maximize the following log-likelihood
\begin{equation} \label{eq:likelihood_weighted_paths}
	\mathcal{L}^{\texttt{ODM}}(f_Y) = \sum_{k=1}^m  \texttt{settled}_k \cdot \log(f_Y(y_k)) + \sum_{k=1}^m (1- \texttt{settled}_k) \cdot \frac{1}{n_{\texttt{path}}} \cdot \sum_{p=1}^{n_{\texttt{path}}} \log(f_Y(y_{k, p})),
\end{equation}
where $f_Y(.)$ is the proposed parametric severity distribution and $\texttt{settled}_k$ is one when claim $k$ settles before the evaluation date $\tau$ and zero otherwise. Contract-specific covariates can be included in the severity distribution $f_Y(.)$. This likelihood includes all possible paths for open claims, whereas traditional severity models average these paths to obtain a best estimate of the ultimate cost of an open claim. Consequently, these traditional methods maximize
\begin{equation} \label{eq:likelihood_best_estimate}
\mathcal{L}^{\texttt{trad}}(f_Y) = \sum_{k=1}^m  \texttt{settled}_k \cdot \log(f_Y(y_k)) + \sum_{k=1}^m (1- \texttt{settled}_k) \cdot  \log \left( f_Y \left( \sum_{p=1}^{n_{\texttt{path}}} \frac{1}{n_{\texttt{path}}} \cdot y_{k, p} \right) \right).
\end{equation}
Our proposed approach for severity modelling (see \eqref{eq:likelihood_weighted_paths}) stays close to traditional pricing practice, but resolves the contradiction between best estimates and actual observations that is present in traditional pricing. We refer to \cite{AlbrecherBladt} for the development of a more general framework to incorporate datapoint uncertainty (e.g.~severity for open claims) into parametric estimation procedures. 

\subsection{Non-life reserving} \label{section:reserving}
Reserving models estimate the aggregated future cost of unsettled, open claims that occurred in past exposure periods. We split the total claims reserve in a reserve for incurred, but not (yet) reported claims, i.e.~the IBNR reserve, and a reserve for reported, but not (yet) settled claims, i.e.~the RBNS reserve. The total reserve, denoted $\mathcal{R}$, is the sum of these two reserve contributions, i.e.~
$$\mathcal{R} = \mathcal{R}^{\texttt{IBNR}} + \mathcal{R}^{\texttt{RBNS}}. $$
We compute the IBNR reserve by aggregating (over all policies $i$) the expected severity for occurred, yet unreported claims, i.e.~
$$ E(\mathcal{R}^{\texttt{IBNR}}) = \sum_{i=1}^n \sum_{j = \tau_i+1}^d \text{E}(N_{i,j}) \cdot \text{E}(Y_i \mid \texttt{rep.delay} = j).$$
Similar to the frequency-severity decomposition in pricing, this formula assumes independence between the number of claims and the claim severity. Estimates for the number of reported claims per year, $N_{ij}$, follow immediately from the occurrence and reporting model proposed in Section~\ref{section:reporting}. Expected claim severity is estimated with the techniques outlined in Section~\ref{section:pricing} for pricing. 

For the RBNS reserve, we compute the future cost of all reported, but not yet settled claims. Hereto, we use the hierarchical reserving model outlined in Section~\ref{section:hierarchical} and simulate the joint evolution of all open claims. As a result of independence assumption (S1), simulating this joint evolution reduces to independently simulating a single path for each open claim. We aggregate the simulated future costs across all claims to obtain an estimate of the total RBNS reserve. A distribution and the expected value of the RBNS reserve are then obtained by repeating these steps. 

\section{Case-studies on pricing and reserving with the ODM} \label{section:casestudy}

\subsection{An insurance portfolio} \label{section:case_insurance}

We first illustrate the occurrence and development model on a European MTPL insurance data set. The portfolio consists of \num{1024805} policies active between January 1, 2007 and December 31, 2016, resulting in \num{78627} reported claims. Policies are restricted to a single calendar year. When policyholders were insured in multiple calendar years, the insured period is broken down by calendar year into multiple records. Table~\ref{table:covariates} lists the available policy and claim covariates.

\begin{table}[ht!]
\centering
\begin{tabular}{p{0.25\linewidth}  p{0.7\linewidth}}
\toprule
\multicolumn{2}{l}{{\bf Policy characteristics}} \\
\midrule
\texttt{Policy ID} & Unique identifier of each policy \\
\texttt{Calendar year} & Calendar year in which this policy was active \\
\texttt{Exposure} & Proportion of the year during which the policyholder was insured \\
\texttt{Young driver} & Indicator (yes/no) whether the policy covers a young driver \\
\texttt{Bonus malus} & Bonus malus level of the driver, ordered from low (best) to high (worst)  \\
\texttt{Fuel} & Fuel type of the vehicle (diesel, gasoline, other, unknown) \\
\midrule
\multicolumn{2}{l}{{\bf Claim development characteristics}} \\
\midrule
\texttt{Policy ID} & Unique identifier of the policy on which the claim is registered \\
\texttt{Claim ID} & Unique identifier of the claim \\
\texttt{Occurrence year} & Calendar year in which the claim occurred \\
\texttt{Reporting year} & Calendar year in which the claim was reported \\
\texttt{Observation year} & Number of calendar years elapsed since reporting the claim \\
\texttt{Settlement} & Indicator (yes/no) whether the claim was settled at the end of the observation year \\
\texttt{Payment} & Amount paid for the claim during the observation year \\
\texttt{Reserve} & Expert opinion on the remaining claim costs after the current observation year \\
\texttt{Incurred} & Sum of amount already paid on a claim and the claim-specific reserve \\
\bottomrule
\end{tabular}
\caption{List of covariates available in the MTPL insurance data. The policy characteristics are registered per policy. For the claim development characteristics one record is registered for each observation year during which the claim was open.}
\label{table:covariates}
\end{table}

\subsubsection{Occurrence and reporting of claims}
In our data set, $96.2\%$ of the observed claims were reported in the year of occurrence and $3.6\%$ in the next year. Only $0.2\%$ of the claims have a reporting delay of more than one calendar year. In this analysis we remove claims with a delay of more than one year to put focus on the bulk of claims reported shortly after occurrence. 

We follow the EM algorithm outlined in Section~\ref{section:reporting} and model in the M-step the occurrence of claims via 
\begin{align}\label{eq:ins_case_occ}
	&N_i \sim \text{POI}(\lambda_i = \texttt{expo}_i \cdot \exp(\beta_{\texttt{cal.year}_i} + \beta_{\texttt{young driver}_i} + \beta_{\texttt{bonus malus}_i} + \beta_{\texttt{fuel}_i})),
\end{align}
and the probability of reporting the claim in its year of occurrence is specified as
\begin{align}\label{eq:ins_case_reporting}
	&p_{i,0} = \logit(\gamma_{\texttt{young driver}_i} + \gamma_{\texttt{bonus malus}_i} + \gamma_{\texttt{fuel}_i}). 
\end{align}

\begin{figure}[ht!]
\centering
\begin{subfigure}{0.9\textwidth}
	\includegraphics[width = \textwidth]{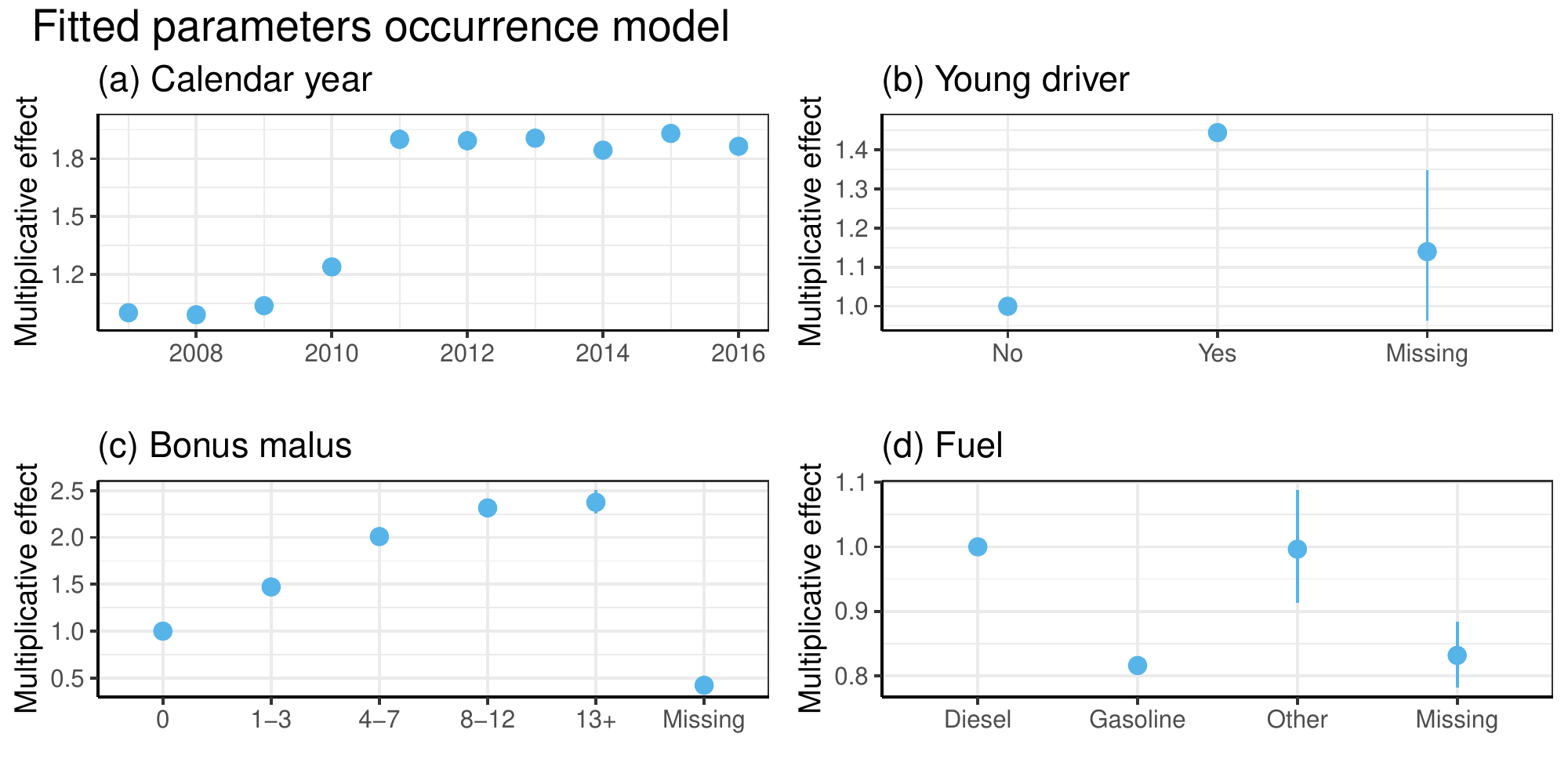}
	\phantomcaption \label{figure:parameters_frequency_calendar_year}
     \phantomcaption \label{figure:parameters_frequency_young_driver}
     \phantomcaption \label{figure:parameters_frequency_bonus_malus}
     \phantomcaption \label{figure:parameters_frequency_fuel}
\end{subfigure}
\begin{subfigure}{0.9\textwidth}
	\includegraphics[width = \textwidth]{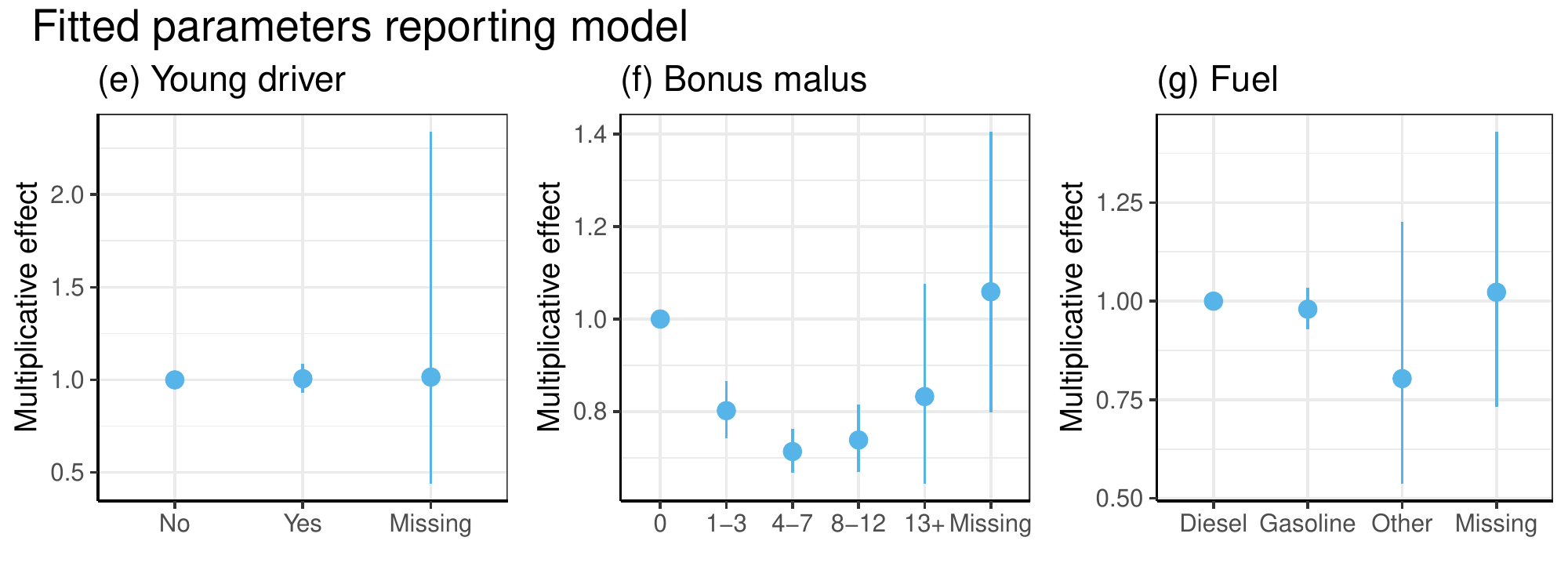}
	\phantomcaption \label{figure:parameters_reporting_young_driver}
    \phantomcaption \label{figure:parameters_reporting_bonus_malus}
    \phantomcaption \label{figure:parameters_reporting_fuel}
\end{subfigure}

\caption{MTPL insurance data set: fitted parameters in the occurrence and reporting models.}
\label{figure:parameters_frequency}
\end{figure}

Figure~\ref{figure:parameters_frequency_calendar_year}, \ref{figure:parameters_frequency_young_driver} \ref{figure:parameters_frequency_bonus_malus} and \ref{figure:parameters_frequency_fuel} show the fitted parameters for the occurrence model specifications in \eqref{eq:ins_case_occ}. Expected claim frequency is higher for young drivers and drivers occupying a higher level in the bonus malus scale. Internal changes insurer cause an administrative increase in the number of registered claims per policy after 2010. This change is captured by the calendar year effect. For fuel, we see that drivers using gasoline have fewer claims. The missing level here essentially corresponds to other motorised vehicles, such as mopeds, being included in the portfolio. Their estimated effect indicates a lower claim risk compared to cars. Figures~\ref{figure:parameters_reporting_young_driver}, \ref{figure:parameters_reporting_bonus_malus} and \ref{figure:parameters_reporting_fuel} display the parameters fitted for \eqref{eq:ins_case_reporting}, the probability of reporting the claim in its year of occurrence. Only bonus malus shows a significant effect with longer reporting delays for policyholders occupying higher bonus malus levels. As a consequence, when directly modelling the claim frequency from the observed claim counts in this MTPL data set, the pricing actuary will underestimate the claim frequency of drivers occupying high bonus malus levels.

\subsubsection{Hierarchical claim development model}\label{sec:hier_dev_ins_case_study}

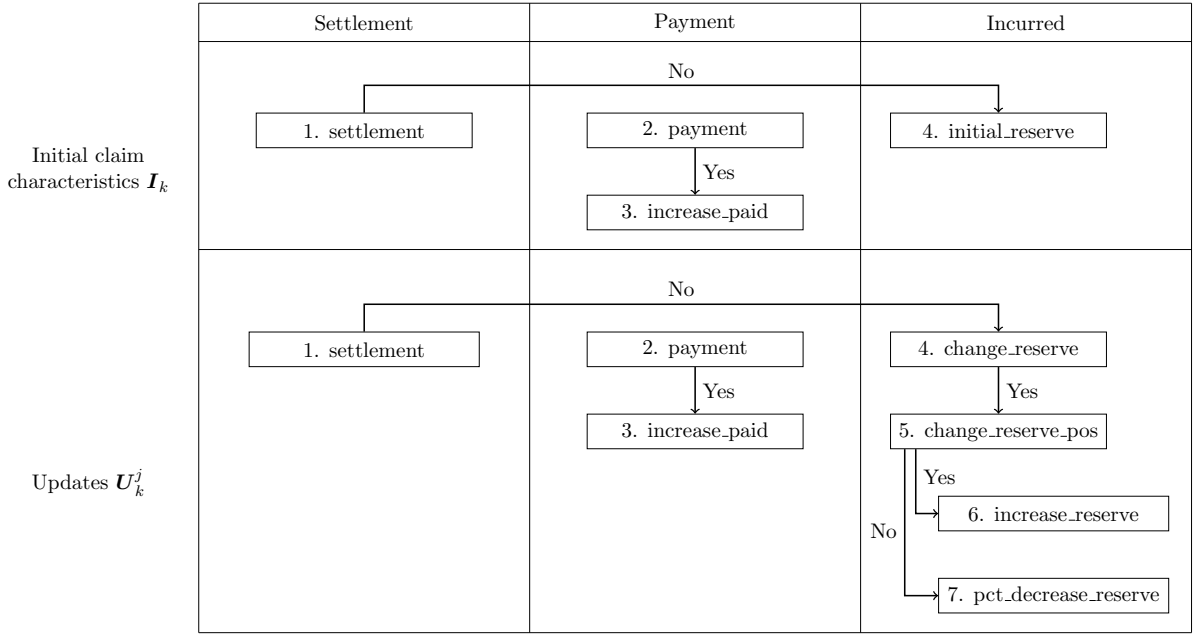
\begin{figure}
\begin{subfigure}[t]{\textwidth}
\centering
\resizebox{\textwidth}{!}{%
\begin{tikzpicture}
	\tikzset{
  		component/.style = {shape=rectangle, draw, minimum width={width("8. pct\_decrease\_reserve")+2pt}, minimum height={18pt}, anchor = north},
  		box/.style = {shape=rectangle, draw, dashed, minimum width={width("8. pct\_decrease\_reserve")+20pt}, minimum height={95pt}, anchor = north},
  		componentwide/.style = {shape=rectangle, draw, minimum width={width("8. pct\_decrease\_reserve")+2pt}, minimum height={18pt}, anchor = north}
	}
	
	\foreach \x in {0, 6, 12, 18}
	{
		\draw(\x, 1.5) -- (\x, -10);
	}
	
	\foreach \y in {1.5, 0.8, -3, -10}
	{
		\draw(0, \y) -- (18, \y);
	}
	
	\node at (3, 1.15) {Settlement};		
	\node at (9, 1.15) {Payment};	
	\node at (15, 1.15) {Incurred};		
	
	\node[rectangle,fill=white] at (-2, -1.55) {\begin{tabular}{c} Initial claim\\characteristics $\boldsymbol{I}_k$ \end{tabular}};

	\node[component] (a1) at (3, -.5) {1. settlement};
	
	\node[component] (a2) at (9, -.5) {2. payment};
	\node[component] (a3) at (9, -2) {3. increase\_paid};	
	
	\node[component] (a5) at (14.5, -.5) {4. initial\_reserve};		
	
	\draw[->,thick] (a2.south) -- (a3.north) node [midway, right] {Yes};
	\draw[->, thick] (a1.north) -- ++(0,0.5) -- ++(11.5, 0) node[midway, above] {No} -| (a5.north)  ;
	
	\node[rectangle,fill=white] at (-2, -7.25) {Updates $\boldsymbol{U}_k^j$};

	\node[component, minimum width={width("pctwdecreasewreserveaw")+2pt}] (e1) at (3, -4.5) {1. settlement};

	\node[component] (e2) at (9, -4.5) {2. payment};
	\node[component] (e3) at (9, -6) {3. increase\_paid};	

	\node[component] (e4) at (14.5, -4.5) {4. change\_reserve};		
	\node[component] (e6) at (14.5, -6) {5. change\_reserve\_pos};		
	\node[component, minimum width={width("pctwdecreasewreserveaw")+2pt}] (e7) at (15.5, -7.5) {6. increase\_reserve};		
	\node[component, minimum width={width("pctwdecreasewreserveaw")+2pt}] (e8) at (15.5, -9) {7. pct\_decrease\_reserve};	
	
	\draw[->, thick] (e6.south)++(-1.5,0) -- ++(0, -0.5) node[right]{Yes} |- (e7.west);	
	\draw[->, thick] (e6.south)++(-1.7,0) -- ++(0, -1.5) node[left]{No} |- (e8.west);
	\draw[->, thick] (e2.south) -- (e3.north) node[midway, right] {Yes};
	\draw[->, thick] (e4.south) -- (e6.north) node[midway, right] {Yes};
	\draw[->, thick] (e1.north) -- ++(0,0.5) -- ++(11.5, 0) node[midway, above] {No} -| (e4.north)  ;
\end{tikzpicture}
}
\end{subfigure}

\caption{MTPL insurance data set: flowchart visualizing the layered structure of the hierarchical claim development model. Solid lines indicate that a layer is modelled conditional on the outcome of a previous layer. Numbers indicate the order in which the layers are modelled.}
\label{figure:flowchart_model_components_insurance}

\end{figure}

\paragraph{The layers.} For each reported claim, the data set tracks the evolution of its settlement status, the amount paid and the amount incurred per observation year since reporting. We use the hierarchical claim development model discussed in Section~\ref{section:hierarchical} to structure the joint evolution of these claim characteristics. Figure~\ref{figure:flowchart_model_components_insurance} sketches its layered structure, tailored to the MTPL insurance data set. A three-layered specification for $\boldsymbol{I}_k$ keeps track of the claim characteristics in the year of reporting. Layer 1 tracks the settlement status of the claim, which is then used as input when modelling whether a payment takes place (layer 2) and (if so) the size of that payment (layer 3). When a claim does not settle in the year of reporting, layer 4 registers the initial reserve set by the claim expert. Beyond the year of reporting a 7-dimensional $\boldsymbol{U}_k^j$ structures the development of claim $k$ in observation year $j-1$ (with $j\geq 2$) since reporting. Hereby, the meaning of layers 1 to 3 does not change. Following a payment, the claim-specific reserve is automatically reduced with the paid amount and upon settlement the incurred is put equal to the paid amount. These are deterministic, automatized operations that do not require any stochastic modelling. However, layer 4 tracks if a (non-automatic) change in the claim-specific reserve takes place (yes or no). Layer 5 then verifies whether that change is positive (yes or no), layer 6 tracks the nominal increase in the reserve (if any) and layer 7 the percentage decrease (if any). A more detailed, technical description of these layers is provided in Appendix~\ref{appendix:layers}.

\paragraph{Predictive model and distributional assumption per layer.} We model each of the layers with a tree-based gradient boosting machine (GBM) \citep{friedman2001}, which additively combines shallow decision trees into one predictor. Three properties make GBMs interesting for automatization. First, automatic binning of continuous covariates allows for capturing non-linear effects. Second, interaction effects are automatically detected when using shallow trees with multiple splits. Third, covariate selection is integrated in the calibration process. For each GBM, we tune five parameters\footnote{We tune the following parameters per considered GBM: number of trees, interaction depth, shrinkage, minimal number of observations per node and bag fraction.} using five-fold cross validation on our training data set. Table~\ref{table:model_specification_distribution_insurance} specifies the distributional assumption for each of the layers in the hierarchical claim development model. We distinguish three types of outcome variables: binary outcomes, percentage changes and numeric outcomes not bounded to the interval $(0, 1)$. We model binary outcomes (e.g.~\texttt{settlement}) with a binomial GBM with logit link function, i.e.~we minimize the loss
$$\mathcal{L}(f^{\texttt{binary}}) = \sum_{i} y_i \cdot f^{\texttt{binary}}(\boldsymbol{z}_i) - \log(1 + \exp(f^{\texttt{binary}}(\boldsymbol{z}_i))),$$
where the sum runs over the available observations for the target layer, $y_i$ is the observed 0/1 outcome, $\boldsymbol{z}_i$ denotes the available covariates for observation $i$ and $f^{\texttt{binary}}(\boldsymbol{z}_i)$ is the prediction delivered by the GBM such that $\text{logit}(P(Y_i=1)) = f^{\texttt{binary}}(\boldsymbol{z}_i)$. Percentage outcomes (e.g.~\texttt{pct\_decrease\_reserve}) are first transformed to the domain $(-\infty, \infty)$ using a logit transform and then modelled using a Gaussian GBM, i.e.~we minimize the loss
$$\mathcal{L}(f^{\texttt{percentage}}) = \sum_{i} (\logit(y_i) - f^{\texttt{percentage}}(\boldsymbol{z}_i))^2.$$
The variance $\sigma^2$ of the Gaussian distribution is estimated as the mean squared error of the residuals, i.e.~
$$\hat{\sigma}^2 = \frac{1}{n} \cdot \sum_{i} (\logit(y_i) - \hat{f}^{\texttt{percentage}}(\boldsymbol{z}_i))^2,$$
where $n$ is the number of observations. Other numeric outcomes (e.g.~\texttt{increase\_paid}) are modelled with a gamma distribution by minimizing the loss 
$$\mathcal{L}(f^{\texttt{numeric}}) = 2 \cdot \sum_{i} \left( \frac{y_i - \exp{(f^{\texttt{numeric}}}(\boldsymbol{z}_i))}{\exp{(f^{\texttt{numeric}}(\boldsymbol{z}_i))}} - \log \frac{y_i}{\exp{(f^{\texttt{numeric}}(\boldsymbol{z}_i))}} \right).$$
The shape parameter $k$ of the gamma distribution is estimated by maximizing the profile likelihood
$$ \mathcal{L}^{shape}(k) = k \cdot \left[ \log\left(\frac{y_i}{\hat{\mu}_i} \right) - \frac{y_i}{\hat{\mu}_i} \right] + \log(k)\cdot k - \log(\Gamma(k)) \quad \text{with} \quad \hat{\mu}_i = \exp{(\hat{f}^{\texttt{numeric}}(\boldsymbol{z}_i)) }.$$

\begin{table}[ht!]
\center
\begin{tabular}{llllll} \toprule
component & distribution & transform & link \\ \midrule
\multicolumn{4}{l}{\bf{Initial claim characteristics} $\boldsymbol{I}$} \\
settlement & binomial &  & logit\\
payment & binomial & . & logit\\
increase\_paid & gamma & . & log\\
initial\_reserve & gamma & . & log  \\
& \\
\multicolumn{4}{l}{\bf{Updates} $\boldsymbol{U}^j$} \\
settlement & binomial & . & logit \\
payment & binomial & . & logit \\
increase\_paid & gamma &  & log \\
change\_reserve & binomial & . & logit \\
change\_reserve\_pos & binomial & . & logit  \\
increase\_reserve & gamma & . & log \\
pct\_decrease\_reserve & Gaussian & logit & . \\ \bottomrule
\end{tabular}

\caption{MTPL insurance data set: distributional specification for the layers in the hierarchical claim development model visualized in Figure~\ref{figure:flowchart_model_components_insurance}.}
\label{table:model_specification_distribution_insurance}
\end{table}

\begin{figure}[ht!]
\centering
\includegraphics[width = \textwidth]{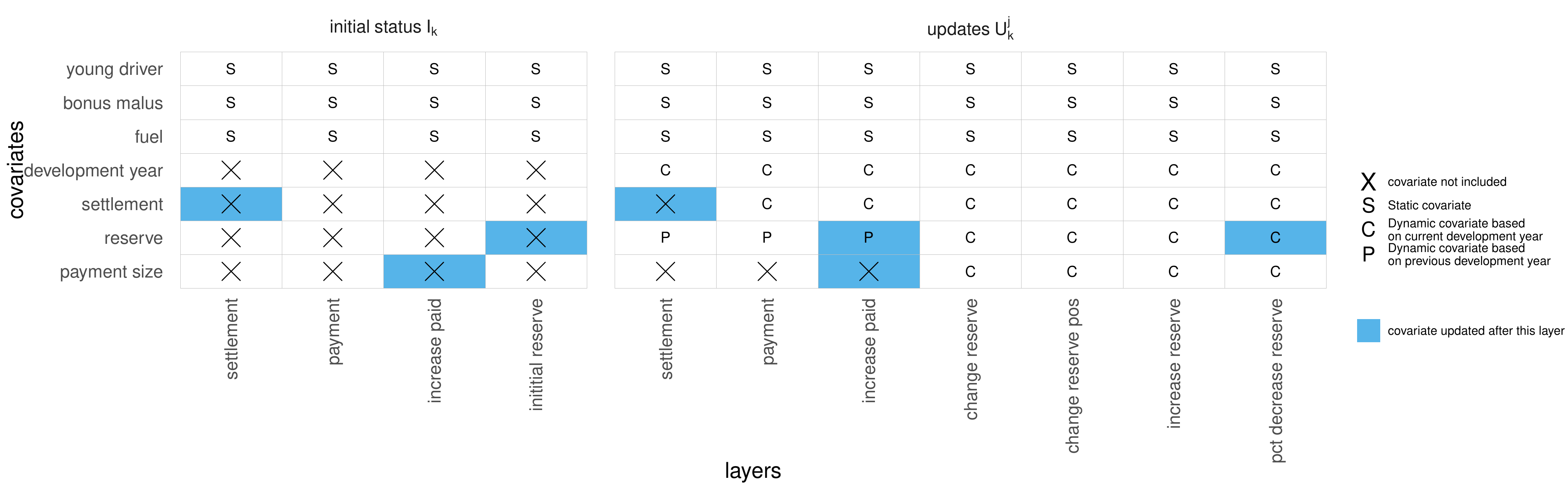}
\caption{MTPL insurance data set: overview of the available covariates for modelling each of the target variables in the layers. The labels and color codes provide information on when each covariate was (re)computed.} \label{figure:covariates_hrm}
\end{figure}

\paragraph{Covariates in the layer-specific predictive model.} We train the layers of the hierarchical reserving model on a data set where each record corresponds to an observation year (since reporting) of a reported claim. Records consist of target variables, static and dynamic covariates. Target variables register the outcome variables of the layers of the hierarchical development model. Static covariates relate to policy characteristics (e.g.,~the fuel type of a car) or claim characteristics (e.g.,~the reporting delay of a claim) and remain constant over the claim development process. Dynamic covariates become available during the claim development process and can be expressed as a function of the target variables. We distinguish three classes of dynamic covariates, namely absolute, relative and aggregated dynamic covariates. Absolute dynamic covariates describe claim characteristics in a fixed, predefined development year (e.g.,~the payment size in development year two since reporting). Once these covariates become available they remain constant for the remainder of the development process. Relative dynamic covariates describe claim characteristics in the current or previous development year (e.g.,~payment size in the previous development year). Aggregated dynamic covariates combine the past claim history in a single aggregated outcome (e.g.,~total amount paid or the current development year). The evolution of these covariates between development years can often be written as a recursive relation. Since we construct one predictive model per layer using data from all development years, we only use relative and aggregated dynamic covariates in our models because these covariates are available independent of the length of the available historical claim information. Figure~\ref{figure:covariates_hrm} summarizes the target variables and the covariates included when modelling these targets in the insurance case-study. For each covariate the figure indicates the covariate type and the layers in which the covariate is updated. 

\subsubsection{Reserving}
\begin{figure}[ht!]
\includegraphics[width = \textwidth]{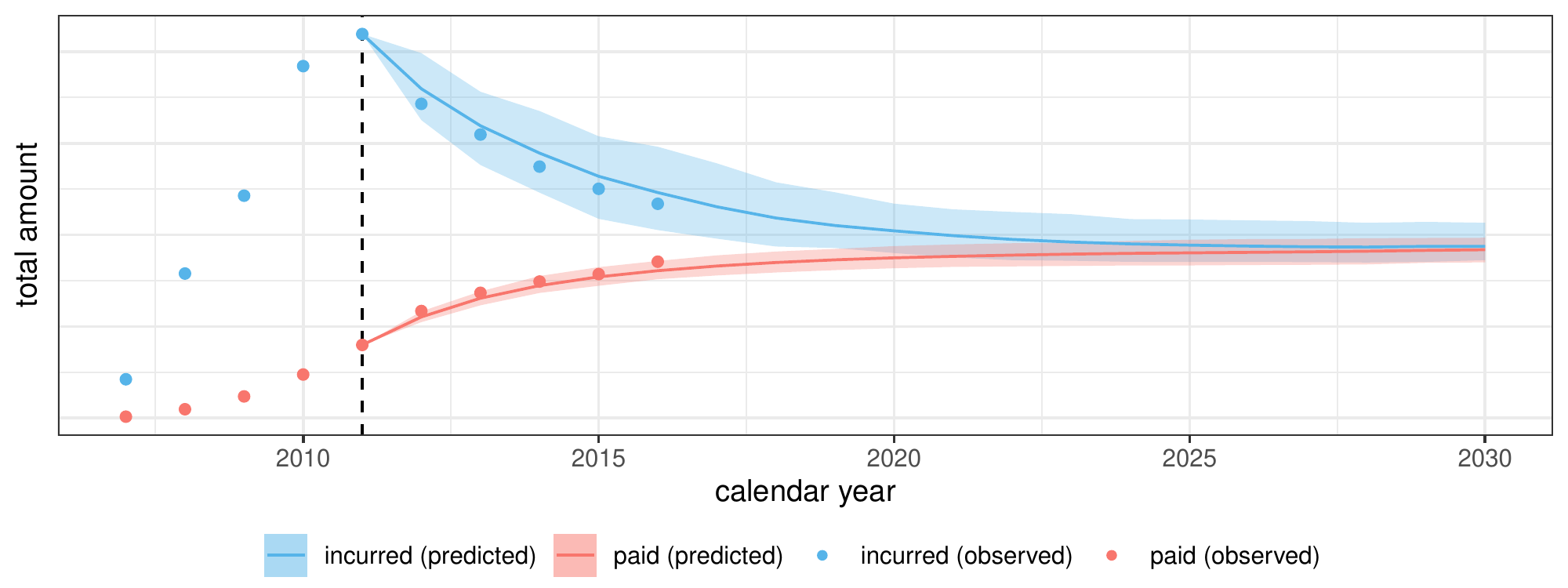}
\caption{MTPL insurance data set: evolution of total paid and incurred amounts for RBNS claims, evaluation date December 31, 2011. 95\% confidence intervals are created from 200 paths sampled for claims reported before the evaluation date and open at the evaluation date. Solid lines indicate expected values and points represent the out-of-time observed amounts for calendar years 2012-2016.} 
\label{fig:ins_case_study_reser}
\end{figure}

\begin{figure}[ht!]
\includegraphics[width = \textwidth]{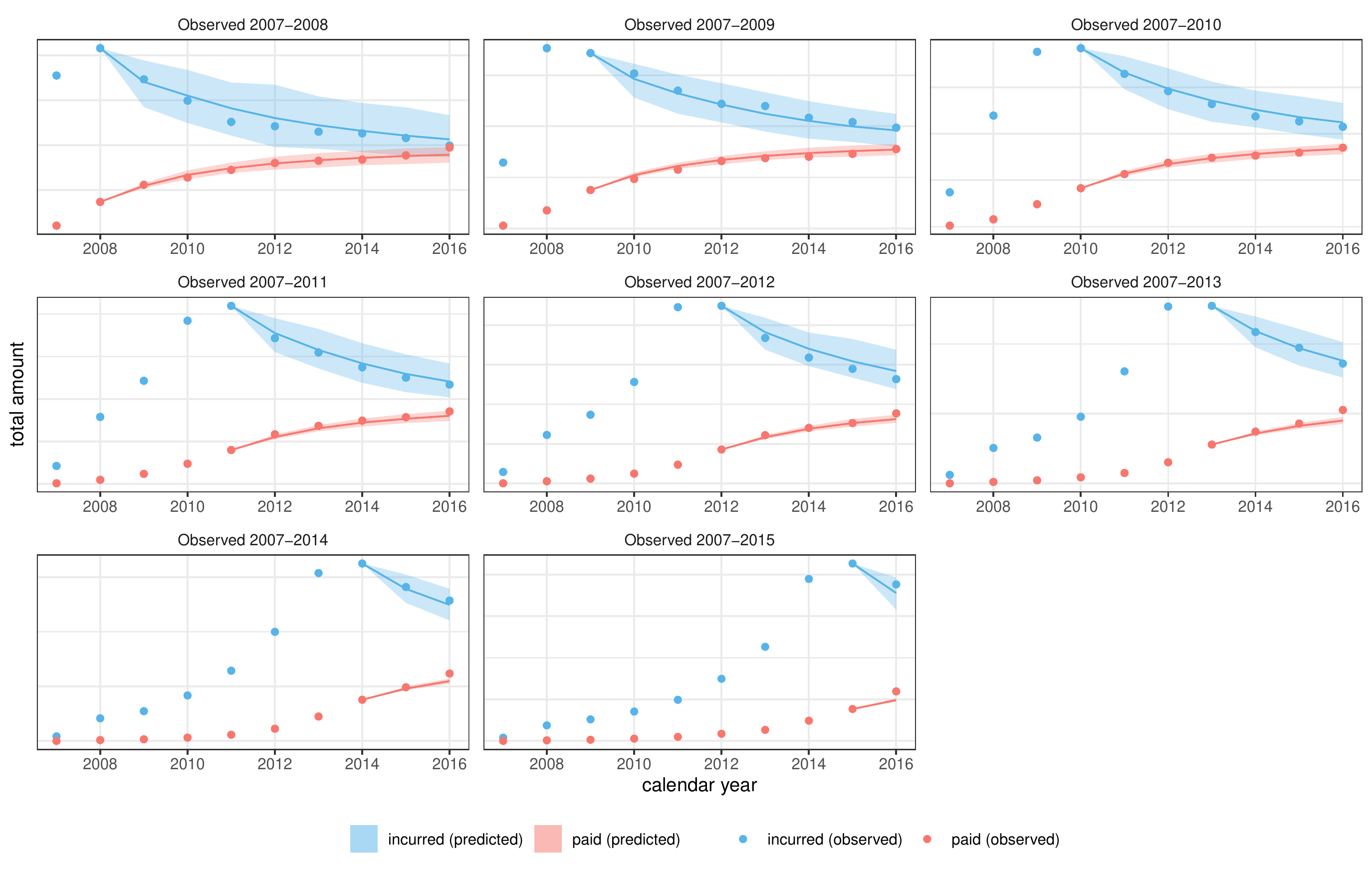}
\caption{MTPL insurance data set: evolution of paid and incurred amounts, when the evaluation date moves from December 31, 2008 to December 31, 2015. 95\% confidence intervals are created from 200 paths sampled for claims reported before the evaluation date and open at the evaluation date. Solid lines indicate expected values and points represent the out-of-time observed amounts.} 
\label{fig:ins_case_study_reser_grid}
\end{figure}

We illustrate the use of the calibrated ODM for reserving. We focus in this illustration on the future development of the RBNS claims. Figure~\ref{fig:ins_case_study_reser} shows 95\% confidence intervals for the evolution of the total amount paid and the total incurred amount for all claims that are open at the end of 2011. Plotted dots indicate the actual observed amounts, as registered in the data set. These realized values fall within the confidence intervals, constructed via 200 simulated paths. The insurer's reserving policy asks claim experts to provide a conservative estimate of the future claim cost. Hence, the decrease in the amount incurred (i.e.,~the paid amount + the outstanding reserve) over time in Figure~\ref{fig:ins_case_study_reser}, the estimate of the claim experts is higher than the actual cost of a claim. The grid of plots in Figure~\ref{fig:ins_case_study_reser_grid} extends this evaluation across multiple evaluation dates. For each of the considered evaluation dates, we estimate the total paid and incurred amounts for claims that are reported before the evaluation date. We then compare the actual realizations with the estimates obtained with our ODM. Our model closely follows the actual portfolio evolution, while claim experts overestimate the total claim cost.

\subsection{A reinsurance portfolio}\label{section:case_reinsurance}

Next, we illustrate our method on a Belgian motor third party liability (MTPL) reinsurance data set registering the detailed development of $\num{4277}$ large motor insurance claims that occurred between 2000 and 2017. These claims originate from 21 underlying MTPL insurance portfolios, which act as the insured clients or policyholders from the reinsurer's perspective. We label these portfolios as \texttt{A}, \texttt{B}, \dots, \texttt{U}. Using our proposed ODM we develop a pricing as well as a reserving strategy for a portfolio of\ excess-of-loss reinsurance contracts. In such an excess-of-loss contract, the reinsurer reimburses the cost of an individual claim exceeding a deductible $D$, up to a limit $L$ \citep{Albrecher2017}.

\begin{figure}[ht!]
\centering
\includegraphics[width = .75\textwidth]{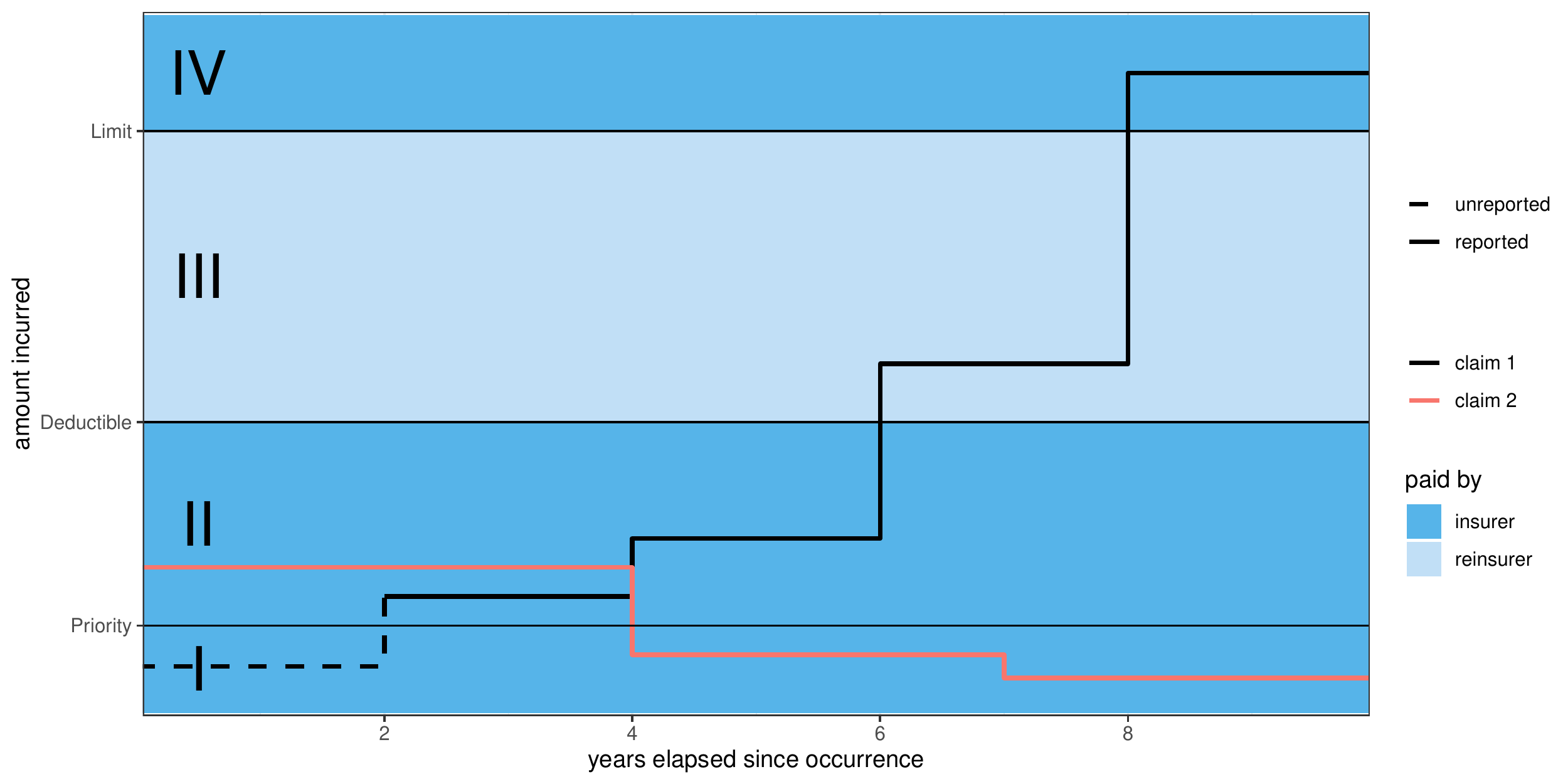}
\caption{MTPL reinsurance data set: illustration of the different thresholds in an excess-of-loss contract. Claims are reported when the incurred first exceeds the reporting  priority. At settlement, the loss between the deductible and the limit is covered by the reinsurer.}
\label{figure:eol_illustration}
\end{figure}

When it comes to large claims, insurers will carefully monitor the evolution of incurred amount, the expected total cost as set by claim handling experts. For the purpose of pricing excess-of-loss reinsurance contracts, insurers are obliged to report a claim to the reinsurer once its incurred exceeds a predefined threshold, the so-called reporting priority. The reporting priority is determined upfront and is specific to both the underlying portfolio and the occurrence year of the claim. Figure~\ref{figure:eol_illustration} visualizes the thresholds (priority, deductible and limit) for the excess-of-loss contract under consideration. In this example, claim 1 (in black) is reported to the reinsurer in year two when its incurred first exceeds the reporting priority. Even when the incurred of claim 2 (in red) falls below the priority in year four, the reinsurer keeps receiving yearly updates on this claim. At settlement, the amount incurred and the paid amount are equal and the reinsurer covers the amount of the loss between the deductible and the limit (region III), while the insurer covers the remaining loss amount (regions I, II and IV).

To evaluate model performance, we split the data and train our model on the years 2000-2014. The remaining years 2015-2017 constitute the out-of-time test data set. Before fitting our ODM we apply three preprocessing steps to the data. First, we remove negative payments. Since the data set contains only a small number of negative payments (accounting for less than $2\%$ of the total amount paid), we believe that the potential gain in model accuracy by incorporating negative payments does not outweigh the implied increase in model complexity. Second, we remove small payments and changes in the incurred of less than 100 euro by combining them with the next large payment or change in the incurred, respectively. Such small changes are frequent, but irrelevant given the large claim sizes in our data. Removing these small changes allows the model to put focus on the important changes in the amount paid and the incurred. Finally, we deflate the payments to the level of 2014 using the inflation curve provided by the reinsurer. After modelling the deflated data, we reinflate the simulated yearly payments to the corresponding payment years when calculating prices and reserves.

\subsubsection{Occurrence and reporting of large claims} \label{section:case_study_reporting}
We slightly adapt Section~\ref{section:reporting} to our reinsurance setting. A policy, indexed with $i$, now refers to a reinsurance contract on an insurance portfolio covering a single underwriting year. In our data set, a claim from policy $i$ is reported when its incurred amount exceeds the reporting priority, denoted $\texttt{priority(i)}$. These priorities are policy-specific, which complicates the comparison of occurrence intensities and reporting delays across policies. Therefore, we choose a new, common priority $P$ shared by all policies. $N_{ij}^{P}$ then denotes the number of claims from policy $i$ for which the incurred first exceeds the priority $P$ in the $(j-1)$-th year since occurrence, i.e.~year $\texttt{occ(i)} + j - 1$. The total number of claims from policy $i$ that exceed the priority $P$ at least once during their development is
$$
N_i^P \coloneqq \sum_{j=1}^{d} N_{ij}^{P}.
$$
Since long reporting delays are common in reinsurance, we set the maximal delay $d$ equal to $15$, the length of the observation window in our data set. The specification of a common reporting priority $P$ naturally restricts the available reinsurance data set to the MTPL insurance portfolios for which $\texttt{priority(i)} \leq P$. Only for these policies we observe the reported claim counts $N_{ij}^{P}$. To investigate the effect of priority $P$ on the estimated price of the excess-of-loss contract, we model the occurrence intensity and reporting delay above three priorities: $\num{750000}$, $\num{1000000}$ and $\num{1250000}$. With these priorities we observe claims from 9, 15 and 15 portfolios (from the original 21), respectively.

Following Section~\ref{section:reporting}, we model the claim occurrence process with a Poisson distribution with intensity
$$ \lambda_{i}:= \lambda(\boldsymbol{x}_i) = e_{i} \cdot \lambda_{\texttt{portfolio(i)}},$$
where $e_{i}$ is the exposure expressed as the number of vehicles insured by policy $i$ and $\lambda_{\texttt{portfolio(i)}}$ denotes the portfolio-specific claim intensity. We model the  reporting probabilities $p_{i, j}$ via their one-to-one connection to the probabilities $q_{i, j}$ introduced in \eqref{eq:reparametrize_q}. The $q$ probabilities are estimated by maximizing the likelihood in \eqref{eq:likelihood_reporting_q} of a binomial GLM with logit link function and
$$q_{i, j} = 1 - \exp(-\exp(\gamma_j + \gamma_{\texttt{portfolio(i)}})),$$
where $\gamma_j$ is the effect of the reporting year and the $\gamma_{\texttt{portfolio(i)}}$ parameters capture reporting delay variations across portfolios.

\begin{figure}[ht!]
\centering
\begin{subfigure}[t]{.9\textwidth}
\centering
\includegraphics[width = \textwidth]{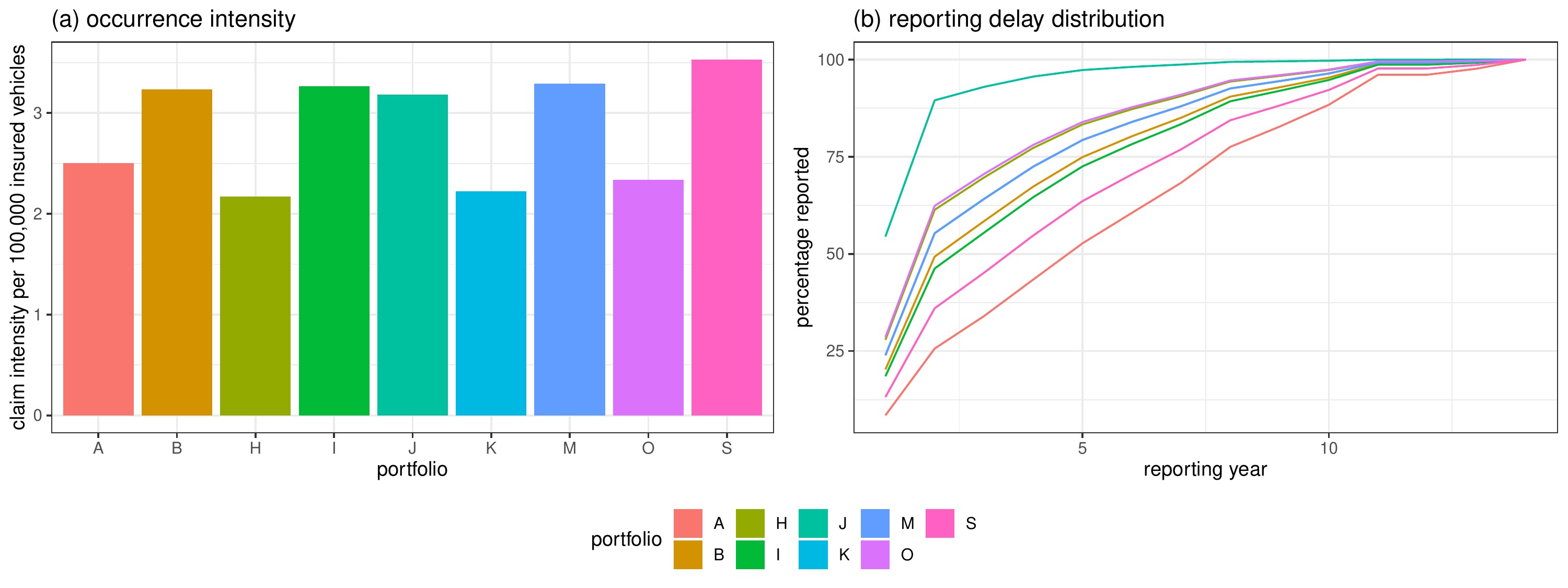}
\phantomcaption \label{figure:fit_occurrence_and_reporting_750k_occ}
\phantomcaption \label{figure:fit_occurrence_and_reporting_750k_rep}
\end{subfigure}
\caption{MTPL reinsurance data set: {\bf (a)} estimated number of claims exceeding the reporting priority of \num{750000} per \num{100000} insured vehicles in 9 insurance portfolios and {\bf (b)} fitted reporting delay distribution per portfolio, where reporting of a claim captures the first exceedance of the incurred claim amount above the priority of \num{750000}.}
\label{figure:fit_occurrence_and_reporting_750k}
\end{figure}

Figure~\ref{figure:fit_occurrence_and_reporting_750k} visualizes the estimated occurrence intensity and the reporting delay distribution when $P = \num{750000}$, using the data from the 9 portfolios available at this priority.  Figure~\ref{figure:fit_occurrence_and_reporting_750k_occ} shows the claim occurrence intensity per $\num{100000}$ insured vehicles for each of these portfolios. We clearly distinguish two regimes in the occurrence intensity: low occurrence intensities ($2.17-2.51$ large claims per $\num{100000}$ vehicles) in insurance portfolios $\texttt{A, H, K}$ and $\texttt{O}$ and high occurrence intensities ($3.18-3.54$ large claims per $\num{100000}$ vehicles) in insurance portfolios $\texttt{B, I, J, M}$ and $\texttt{S}$. This split in two regimes could indicate a different share of more exposed vehicles (e.g.,~buses and trucks) insured in these portfolios.

Figure~\ref{figure:fit_occurrence_and_reporting_750k_rep} shows the estimated reporting delay distribution per insurance portfolio. The incurred amount is volatile in the first years after the occurrence of claims, when there is significant uncertainty regarding the final claim amount e.g.,~because the physical damage has not yet been decided in court or the victim has not yet reached the age of majority. As a result, a portfolio of reinsurance contracts is characterized by long reporting delays between the occurrence of a claim and the moment its incurred first exceeds the reporting priority. Moreover, since each insurer follows its own reserving policy, we find considerable differences in reporting delay.

The insights revealed in Figure~\ref{figure:fit_occurrence_and_reporting_750k} are important for reinsurers when pricing reinsurance contracts issued to these insurance portfolios. Reinsurers can share these insights with their policyholders, i.e.~the insurers. This enables insurers to benchmark the observed reporting delay for their portfolio to the market and provides incentives to insurers with long reporting delays to put more focus on accurately reserving large claims.

\subsubsection{A hierarchical model for the development of large claims after reporting} \label{section:casestudy_claim_development}

For each reported claim, our data set tracks the evolution of the settlement status, the amount paid and the amount incurred per year. Since these events in a claim's development process are clearly dependent (e.g.,~no payments after settlement, low settlement probability when the outstanding reserve is large), we use the hierarchical model of Section~\ref{section:hierarchical} to model the joint evolution of these claim characteristics.

\begin{figure}
\begin{subfigure}[t]{\textwidth}
\centering
\resizebox{\textwidth}{!}{%
\begin{tikzpicture}
	\tikzset{
  		component/.style = {shape=rectangle, draw, minimum width={width("8. pct\_decrease\_reserve")+2pt}, minimum height={18pt}, anchor = north},
  		box/.style = {shape=rectangle, draw, dashed, minimum width={width("8. pct\_decrease\_reserve")+20pt}, minimum height={95pt}, anchor = north},
  		componentwide/.style = {shape=rectangle, draw, minimum width={width("8. pct\_decrease\_reserve")+2pt}, minimum height={18pt}, anchor = north}
	}
	
	\foreach \x in {0, 6, 12, 18}
	{
		\draw(\x, .6) -- (\x, -11.5);
	}
	
	\foreach \y in {-.1, .6, -3, -11.5}
	{
		\draw(0, \y) -- (18, \y);
	}
	
	\node at (3, .25) {Settlement};		
	\node at (9, .25) {Payment};	
	\node at (15, .25) {Incurred};		
	
	\node[rectangle,fill=white] at (-2, -1.55) {\begin{tabular}{c} Initial claim\\characteristics $\boldsymbol{I}_k$ \end{tabular}};		

	\node[component] (a2) at (9, -.5) {2. payment};
	\node[component] (a3) at (9, -2) {3. pct\_paid};	
	
	\node[component] (a5) at (14.5, -.5) {1. excess\_incurred};		
	
	\draw[->,thick] (a2.south) -- (a3.north) node [midway, right] {Yes};

	\node[rectangle,fill=white] at (-2, -7.25) {Updates $\boldsymbol{U}_k^j$};

	\node[component, minimum width={width("pctwdecreasewreserveaw")+2pt}] (e1) at (3, -4.5) {1. settlement};

	\node[component] (e2) at (9, -4.5) {2. payment};
	\node[component] (e3) at (9, -6) {3. increase\_paid};	

	\node[component] (e4) at (14.5, -4.5) {4. change\_reserve};		
	\node[component] (e5) at (14.5, -6) {5. reserve\_is\_zero};		
	\node[component] (e6) at (14.5, -7.5) {6. change\_reserve\_pos};		
	\node[component, minimum width={width("pctwdecreasewreserveaw")+2pt}] (e7) at (15.5, -9) {7. increase\_reserve};		
	\node[component, minimum width={width("pctwdecreasewreserveaw")+2pt}] (e8) at (15.5, -10.5) {8. pct\_decrease\_reserve};	
	
	\draw[->, thick] (e6.south)++(-1.5,0) -- ++(0, -0.5) node[right]{Yes} |- (e7.west);	
	\draw[->, thick] (e6.south)++(-1.7,0) -- ++(0, -1.5) node[left]{No} |- (e8.west);
	\draw[->, thick] (e2.south) -- (e3.north) node[midway, right] {Yes};
	\draw[->, thick] (e4.south) -- (e5.north) node[midway, right] {Yes};
	\draw[->, thick] (e5.south) -- (e6.north) node[midway, right] {No};
	\draw[->, thick] (e1.north) -- ++(0,0.5) -- ++(11.5, 0) node[midway, above] {No} -| (e4.north)  ;
\end{tikzpicture}
}
\phantomcaption \label{figure:flowchart_model_components_initial}
\phantomcaption \label{figure:flowchart_model_components_update}
\end{subfigure}

\caption{MTPL reinsurance data set: flowchart visualizing the layered structure of the hierarchical claim development model. Solid lines indicate that a layer is modelled conditional on the outcome of a previous layer. Numbers indicate the order in which the layers are modelled.}
\label{figure:flowchart_model_components}

\end{figure}
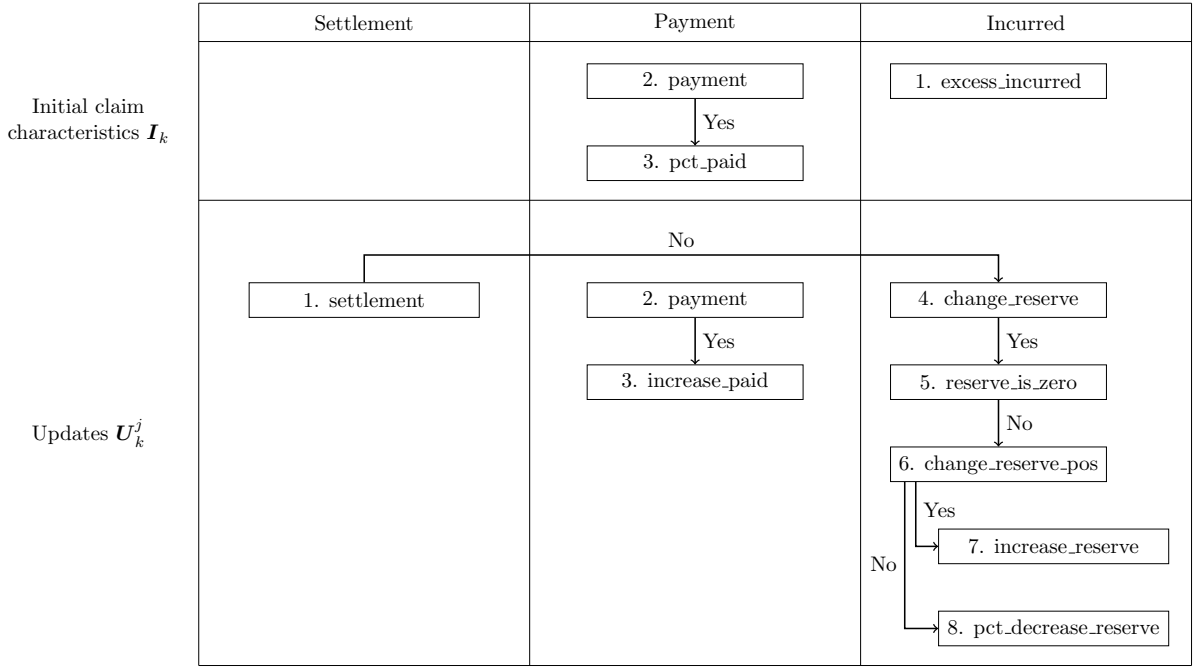

\paragraph{The layers.} We choose a reporting priority, $P = \num{750000}$, and interpret $\boldsymbol{I}_k$ as the dynamic claim characteristics registered for claim $k$ when its incurred amount first exceeds $\num{750000}$. The top row in Figure~\ref{figure:flowchart_model_components} visualizes the 3-layer hierarchical structure for $\boldsymbol{I}_k$, a three-dimensional vector. At reporting, the incurred exceeds the reporting priority of $\num{750000}$. Layer 1 (the first entry in the vector $\boldsymbol{I}_k$) captures the excess amount of the incurred above this reporting priority, i.e.~the difference between the initial incurred amount and this reporting priority. As a result of the data preprocessing step, we only record differences of at least 100 euro. The outcome of this first layer is an input when modelling the amount paid in layer 2 and 3, the second and third entries in $\boldsymbol{I}_k$. Layer 2 tracks whether a part of the incurred is paid at reporting (yes or no). In case of a payment, layer 3 stores the amount paid at reporting as a percentage of the total incurred. We do not model the settlement status in the year of reporting, because claims never settle immediately at reporting in this reinsurance data set. The bottom row in Figure~\ref{figure:flowchart_model_components} visualizes the 8-layer hierarchical specification for the update vector $\boldsymbol{U}_k^j$ in the $(j-1)$-th year since reporting (with $j\geq 2$). First, layer 1 registers the settlement status of a claim. Settlement status is used as an input when modelling payments and changes in the incurred. Layer 2 tracks the presence of a payment and layer 3 captures the size of a payment conditional on the presence of a payment. Note that we only take payments above 100 euro into account. Following a payment, we deterministically decrease the claim-specific reserve by the payment size. When the claim settles, the incurred is set equal to the total amount paid. This is a deterministic operation and no modelling is required. However, when a claim does not settle, layers 4 to 8 express the reserve changes. These five layers let our model capture a drop of the reserve to zero, a nominal increase in the reserve or a decrease expressed as a percentage of the outstanding reserve. A more detailed, technical description of the 11 (3+8) layers is available in Appendix~\ref{appendix:layers}. 

\paragraph{Predictive model and distributional assumption per layer.} Similar to the insurance case-study, we model each of the layers with a tree-based gradient boosting machine (GBM). Table~\ref{table:model_specification_distribution} specifies the distributional assumption per layer. We distinguish three types of outcome variables: binary outcomes, percentage changes and numeric outcomes not bounded to the interval $(0, 1)$. For the binary outcomes and the percentage changes we follow the distributional assumptions discussed in Section~\ref{sec:hier_dev_ins_case_study}. However, other numeric outcomes (e.g.~\texttt{increase\_paid}) are in this example left-truncated at 100 because of the removal of small payments and changes in the incurred in the data preprocessing step. Moreover, these outcomes are heavily right skewed given our reinsurance context. Therefore, we first normalize these observations by applying a power transform, i.e.~we replace the random variable $X$ by $X^p$ for some power $p$, and then estimate a truncated Gaussian GBM for the normalized outcomes. We minimize the following loss function 
\begin{equation}  \label{eq:likelihood_power_a}
\resizebox{.94\hsize}{!}{$\mathcal{L}(f^{\texttt{numeric}}, \sigma, p) = \sum_{i} \left(\log(\sigma) + \frac{(y_i^p - f^{\texttt{numeric}}(\boldsymbol{z}_i) )^2}{2\sigma^2} + \log(\Phi(100^p \mid f^{\texttt{numeric}}(\boldsymbol{z}_i), \sigma)) - \log(p) - p \cdot \log(y_i)\right), $}
\end{equation}
where $p$ is the exponent in the power transform and $\Phi(\cdot \mid \mu, \sigma)$ is the cdf of the Gaussian distribution with mean $\mu$ and standard deviation $\sigma$. We opt for a two step calibration approach. First, we minimize $\eqref{eq:likelihood_power_a}$ with respect to $\sigma$, $p$ and a constant $f^{\texttt{numeric}}(\cdot)$. Second, we re-estimate $f^{\texttt{numeric}}(\cdot)$ and $\sigma$ using a truncated Gaussian GBM, while keeping the power $p$ fixed.

\begin{table}[ht!]
\center
\begin{tabular}{llllll} \toprule
component & distribution & transform & link \\ \midrule
\multicolumn{4}{l}{\bf{Initial claim characteristics} $\boldsymbol{I}$} \\
excess\_incurred & trunc.Gaussian & $(.)^{0.117}$ & .\\
payment & binomial & . & logit\\
pct\_paid &Gaussian & logit & .  \\
& \\
\multicolumn{4}{l}{\bf{Updates} $\boldsymbol{U}^j$} \\
settlement & binomial & . \\
payment & binomial & . & logit \\
increase\_paid & trunc.Gaussian & $(.)^{0.155}$ & . \\
change\_reserve & binomial & . & logit \\
reserve\_is\_zero & binomial & . & logit  \\
change\_reserve\_pos & binomial & . & logit  \\
increase\_reserve & trunc.Gaussian & $(.)^{0.105}$ & . \\
pct\_decrease\_reserve & Gaussian & logit & . \\ \bottomrule
\end{tabular}

\caption{MTPL reinsurance data set: distributional specification for the model components in the hierarchical claim development model visualized in Figure~\ref{figure:flowchart_model_components}. The exponents of the power transformations were calibrated based on the available training data.}
\label{table:model_specification_distribution}
\end{table}

\begin{figure}[ht!]
\centering
\includegraphics[width = \textwidth]{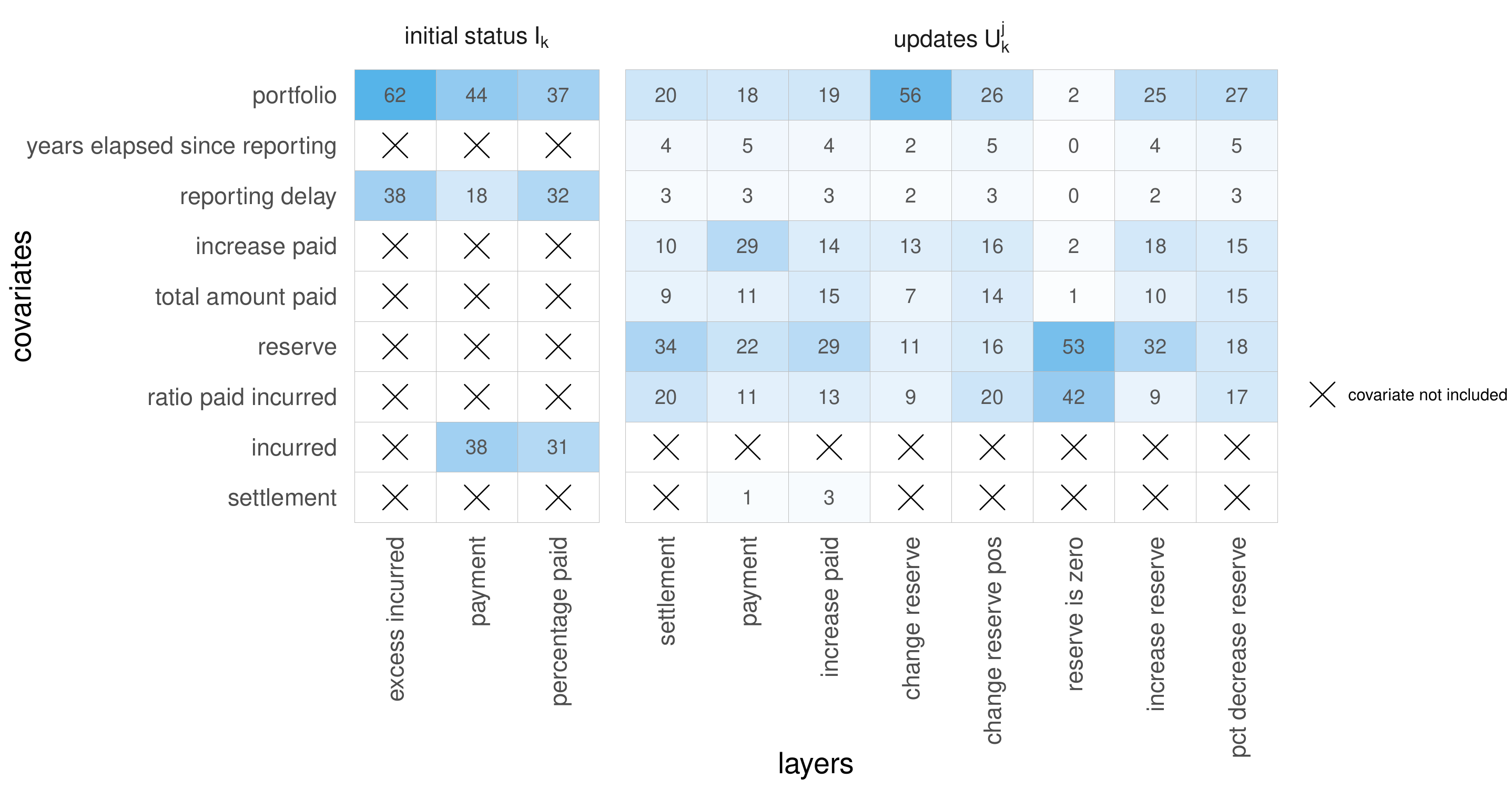}
\caption{MTPL reinsurance data set: available covariates and their relative importance on the layers of the hierarchical model for the initial status $\boldsymbol{I}_k$ and the updates $\boldsymbol{U}_k^j$. Relative importance is computed as the decrease in the loss function captured by the tuned GBM over all splits using a specific covariate, relative to the total decrease in the loss function caused by all covariates. The relative importance is evaluated on the training data.}
\label{figure:variable_importance_gbm}
\end{figure}

\paragraph{Feature effects.} Figure~\ref{figure:variable_importance_gbm} shows for each fitted GBM the relative importance of the included covariates. The variable importance of a covariate is the decrease in the GBM's loss function over all tree splits using the covariate under consideration, when optimal values are used for the tuning parameters. \texttt{portfolio} is an important covariate in almost all layers, indicating clear differences in the handling of large claims between the insurers in the data set. Most noteworthy is the effect of the portfolio on the layer \texttt{change\_reserve}. In some portfolios experts re-evaluate their large claims almost every year, whereas other insurers rarely update their large claims. Both covariates \texttt{reserve} (i.e. incurred - paid) and \texttt{ratio paid incurred} (i.e. $\frac{\text{paid}}{\text{incurred}}$) describe a relationship between the incurred amount and the paid amount. Together these covariates are for many layers the most important predictors for a claim's future development. In traditional, aggregated reserving models, claim development depends only on the number of \texttt{years elapsed since reporting}, i.e.~the development year. Surprisingly, this covariate becomes irrelevant when more informative claim characteristics are available.

\begin{figure}[ht!]
\centering
\begin{subfigure}[t]{.9\textwidth}
\centering
\includegraphics[width = \textwidth]{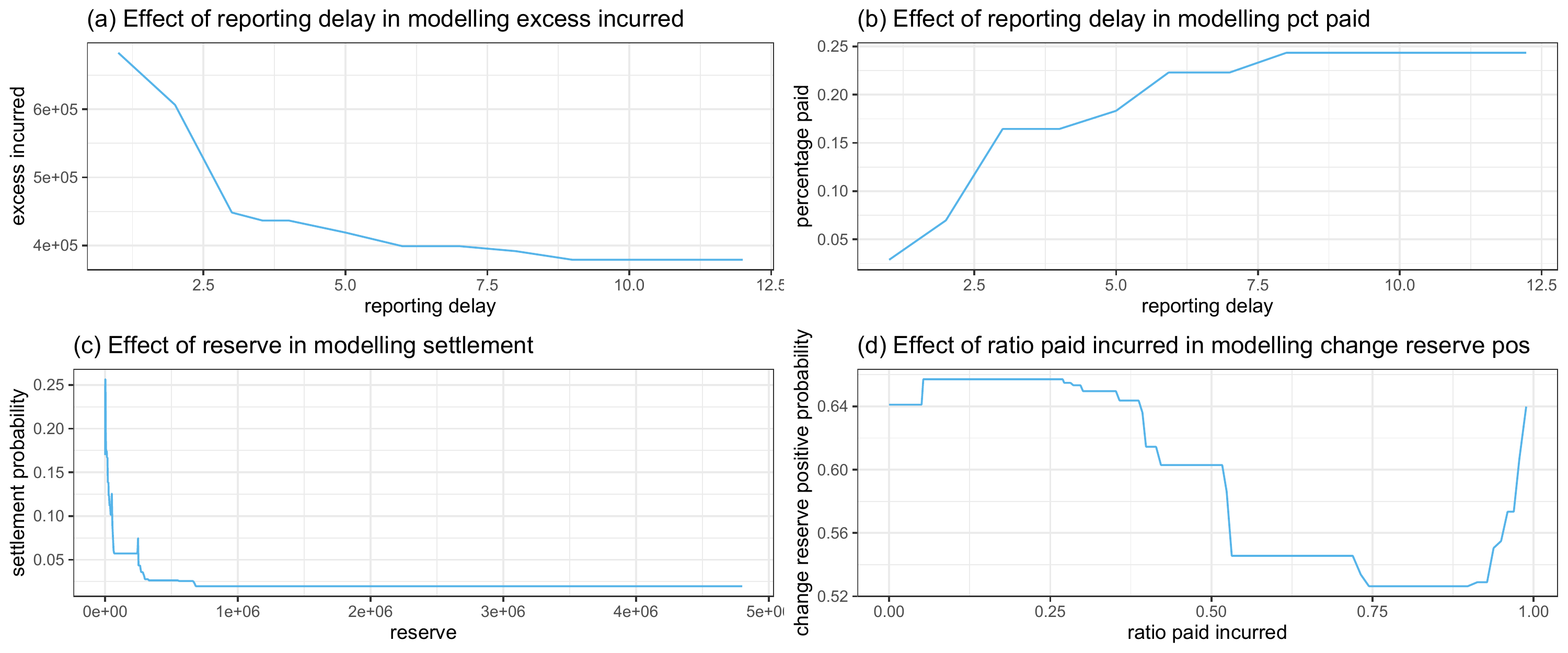}
\phantomcaption \label{figure:pdp_incurred_delay}
\phantomcaption \label{figure:pdp_pct_paid_delay}
\phantomcaption \label{figure:pdp_settlement_reserve}
\phantomcaption \label{figure:pdp_change_reserve_pos_ratio}
\end{subfigure}
\caption{MTPL reinsurance data set: selection of partial dependence plots in the hierarchical claim development model.}
\label{figure:claim_development_pdp}
\end{figure}

Figure~\ref{figure:claim_development_pdp} uses partial dependence plots to visualize the marginal effect of selected covariates on the outcome layers in the hierarchical model. The \texttt{excess incurred} is smaller for claims reported to the reinsurer after a long delay (Figure~\ref{figure:pdp_incurred_delay}) and for these claims a larger fraction of the incurred has already been paid at reporting by the insurer (Figure~\ref{figure:pdp_pct_paid_delay}). This is intuitive taking into account that the reporting delay is different from the insurer's and the reinsurer's perspective and that large claims are often quickly reported to the insurer. Late reporting of a claim to the reinsurer thus gives the insurer more time to make claim payments. As expected, Figure~\ref{figure:pdp_settlement_reserve} shows that claims are likely to settle when the outstanding reserve is near zero. The incurred is more likely to increase when either little has been paid yet for the claim or when the paid amount is close to the incurred amount (Figure~\ref{figure:pdp_change_reserve_pos_ratio}).

\subsubsection{Pricing an excess-of-loss reinsurance contract} \label{section:cs_pricing}
We price an excess-of-loss reinsurance contract covering losses from individual claims exceeding a deductible $D = \num{2500000}$ up to a limit $L = \num{5000000}$. Following the frequency-severity decomposition, the pure premium $\pi^{P}$ is
\begin{eqnarray*}
\pi^P = E(N^P) \cdot E(((Y^P \wedge L) - D)_+).
\end{eqnarray*}
Here $N^P$ and $Y^P$ are the frequency and severity, respectively, of claims reported above a priority $P$, $(Y^P \wedge L)$ denotes the minimum of $Y^P$ and $L$, and $(Y-D)_+$ equals $Y-D$ if $Y \geq D$ and zero otherwise. The pure reinsurance premium scales directly with the number of insured vehicles in the underlying insurance portfolio, i.e.~the exposure $e_i$. In this section we set the exposure to one and hence compute the premium for a single insured vehicle.

In Section \ref{section:case_study_reporting}, we calibrated the occurrence of claims from policy $i$ as
$$N_i^P \sim \texttt{Poisson}(e_i \cdot \lambda^P_{\texttt{portfolio(i)}}),$$
where $\lambda^P_{\texttt{portfolio(i)}}$ is the expected number of clams per insured vehicle from $\texttt{portfolio}(i)$ exceeding the reporting priority $P$, i.e.~our frequency estimate. Figure \ref{figure:fit_occurrence_and_reporting_750k_occ} pictures the fitted parameters $\lambda^P_{\texttt{portfolio(i)}}$ for the various portfolios in our data set when using a reporting priority of $\num{750000}$.

Section~\ref{section:pricing} outlined two strategies for simulating the claim severity distribution. The first strategy simulates a large number of paths for a new claim from ground up, whereas the second strategy simulates the future development of open claims. We illustrate the use of both simulation strategies to model the severity distribution of a new claim from portfolio \texttt{A} when the reporting priority $P$ is equal to $\num{750000}$. Appendix~\ref{sec:sim_strat_sev} outlines the details of both simulation strategies.

\begin{figure}[ht!]
\centering
\begin{subfigure}[t]{.9\textwidth}
\centering
\includegraphics[width = \textwidth]{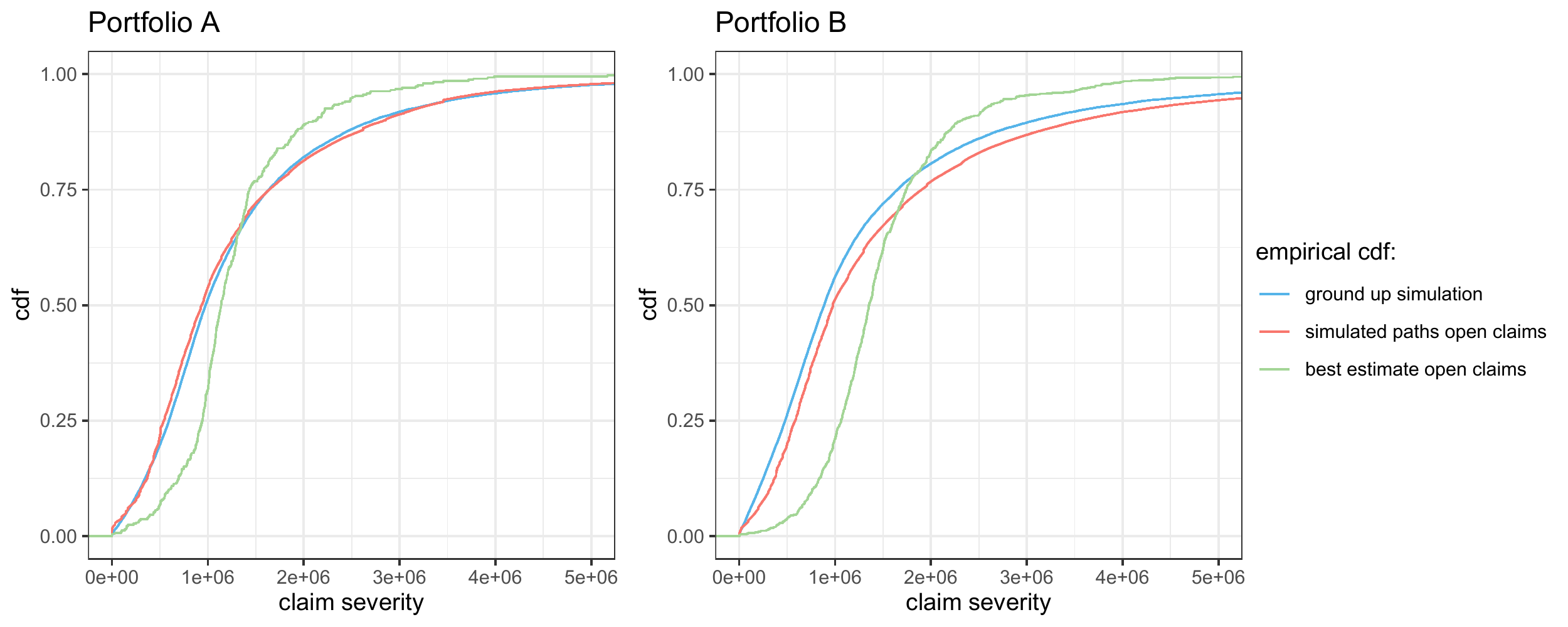}
\end{subfigure}
\caption{MTPL reinsurance data set: simulated severity distribution for claims from portfolio $\texttt{A}$ and $\texttt{B}$ above a reporting priority of $\num{750000}$. For each portfolio, we show the ecdf of the severity distribution based on $\num{20000}$ from ground up simulated new claims (blue), observed claims complemented with 200 simulated paths per open claim (red) and observed claims where open claims have been replaced by best estimates (green).}
\label{figure:empirical_cdf_claim}
\end{figure}

\paragraph{Comparing simulated severity distributions} Figure~\ref{figure:empirical_cdf_claim} shows the empirical claim severity distributions based on simulated paths from ground up (blue) and simulated paths for the future development of observed claims (red). Since we price an excess-of-loss contract with a limit $L$ of $\num{5000000}$, we only show the distribution of the ultimate claim severity below $\num{5000000}$. For portfolio $\texttt{A}$ both simulation strategies result in nearly identical severity distributions. Repeating the same approach for portfolio $\texttt{B}$, we retrieve a more heavy tailed severity distribution when simulating paths for the future development of observed claims. Figure~\ref{figure:empirical_cdf_claim} compares the claim severity distributions proposed in our paper with the empirical cdf based on best estimates (green), where for each open claim the best estimate is calculated by averaging the ultimate claim severity over the 200 simulated paths. This distribution has the same mean, but a lower variance than the distribution based on the simulated paths for observed claims. As argued in Section~\ref{section:claim_development_model}, an underestimation of the variance can have severe implications when pricing complex (re)insurance products. 

\begin{figure}[ht!]
\centering
\begin{subfigure}[t]{.9\textwidth}
\centering
\includegraphics[width = \textwidth]{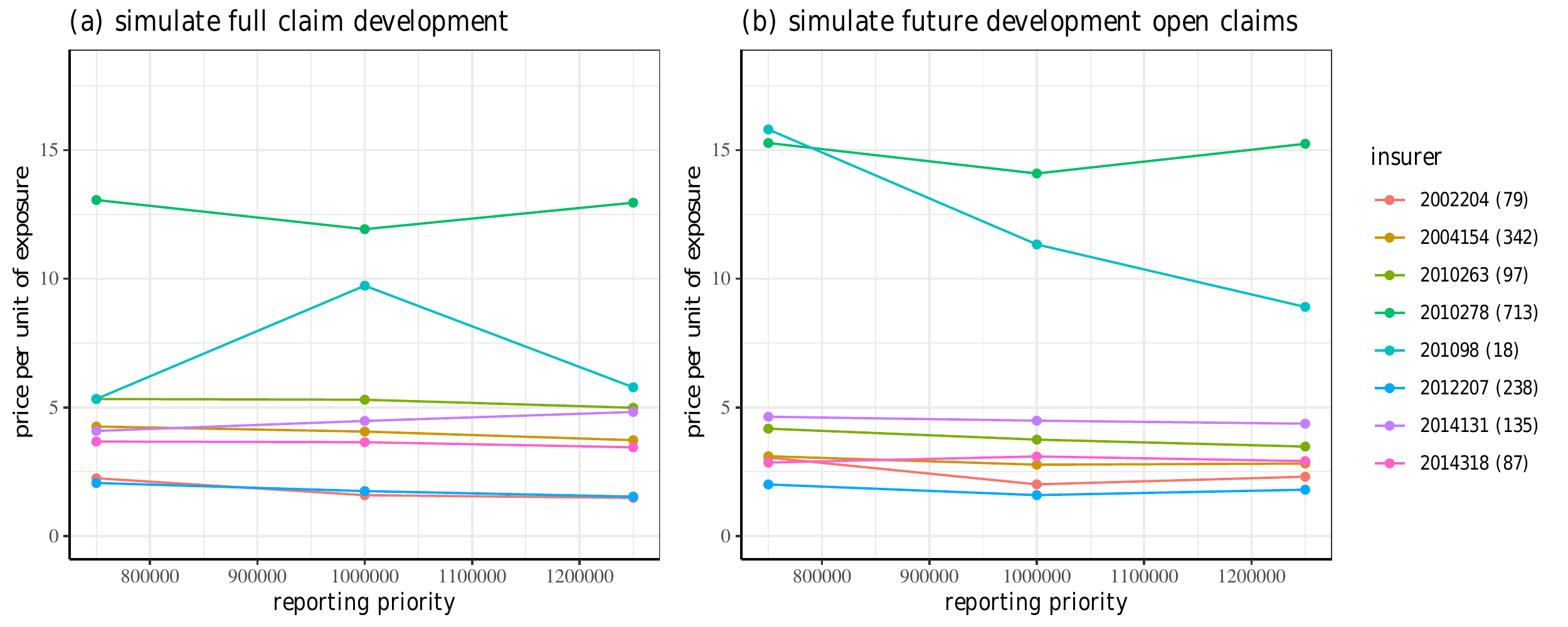}
\phantomcaption \label{figure:price_reporting_priority_approach1}
\phantomcaption \label{figure:price_reporting_priority_approach2}
\end{subfigure}
\caption{MTPL reinsurance data set: estimated cost per insured vehicle for an excess-of-loss contract with deductible $\num{2500000}$ and limit $\num{5000000}$. Claim severity is estimated based on {\bf(a)} simulating $\num{20000}$ new claims from ground up and {\bf(b)} observed claims complemented with $200$ simulated paths per open claim. Prices are computed at reporting priorities: $\num{750000}$, $\num{1000000}$ and $\num{1250000}$. }
\label{figure:price_reporting_priority}
\end{figure}

\paragraph{Pricing an excess-of-loss contract}
Across the available portfolios and for three chosen reporting priorities, Figure~\ref{figure:price_reporting_priority} shows the pure premium per insured vehicle for an excess-of-loss policy. Hereby we use the severity distribution obtained via (lhs) simulating $\num{20000}$ new claims from ground up and (rhs) observed claims complemented with $200$ simulated paths for the future development of each open claim. In theory, the choice of the reporting priority should not influence the price of the reinsurance contract. In practice however some differences in the estimated pure premium arise since the priority determines the available historical claims when calibrating the ODM. We investigate the sensitivity of the pure premium with respect to the priority by modelling the frequency and severity above a reporting priority of $\num{750000}$, $\num{1000000}$ and $\num{1250000}$. For most portfolios, the price remains relatively constant when changing the priority, but larger variations are observed for some small portfolios (e.g.,~portfolio \texttt{S}). These variations mainly result from the claim frequency model for which the priority determines the available claims when training the model. Since we detected two regimes in the occurrence intensity in Figure~\ref{figure:fit_occurrence_and_reporting_750k_occ}, our frequency model could be made more robust by estimating a single occurrence intensity parameter per regime. Estimated prices are comparable when (lhs) simulating new claims from ground up and (rhs) simulating paths for open claims. Price differences are often the result of realised extreme claims, which more heavily influence the estimated cost based on the paths generated for the observed claims.

\subsubsection{Reserving for reinsurance contracts}
Reserving actuaries estimate the aggregated, future cost for claims from past exposure years. In reinsurance, these costs depend on the structure of the contract sold. We estimate the reserve that should be held by the reinsurer under two contract types. The first type of contract covers all losses booked on claims for which the incurred exceeds the reporting priority of $\num{750000}$ at least once during the claim's development. Although this contract is not sold in practice, it is relevant to be considered because of its similarities with the reserving problem in the classical insurance setting. For accurately reserving this contract, the ODM should capture the average development pattern of claims over time sufficiently well. The second type of reinsurance contract covers the loss of an individual claim in our reinsurance data set between the deductible of $\num{2500000}$ and the policy limit of $\num{5000000}$, i.e.~the reinsurance contract that we priced in Section~\ref{section:cs_pricing}. This contract puts focus on the performance of our ODM for large claims. For convenience, we assume that the contracts under consideration cover claims from occurrence years 2000-2014 on the available nine MTPL insurance portfolios with a reporting priority below $\num{750000}$.

Reserving follows the outline explained in Section~\ref{section:reserving} and relies on the proposed calibration strategies for pricing as discussed in Section~\ref{section:cs_pricing}. For the IBNR reserve, we predict the number of occurred, yet unreported claims and their expected reporting date from the occurrence and reporting model. We estimate the severity of these unreported claims by simulating new claims from ground up. In these simulations we account for the effect of long reporting delays on the reinsurance claim development process (Figure~\ref{figure:pdp_incurred_delay} and \ref{figure:pdp_pct_paid_delay}). For the RBNS reserve, we use the ODM to simulate the future development of the reported, open claims.

\paragraph{Reserving for a portfolio of reinsurance contracts that cover ground-up losses}

\begin{figure}[ht!]
\centering
\begin{subfigure}[t]{.9\textwidth}
\phantomcaption \label{figure:reserve_ground_up_total}
\phantomcaption \label{figure:reserve_ground_up_ibnr}
\phantomcaption \label{figure:reserve_ground_up_rbns}
\end{subfigure}
\includegraphics[width = \textwidth]{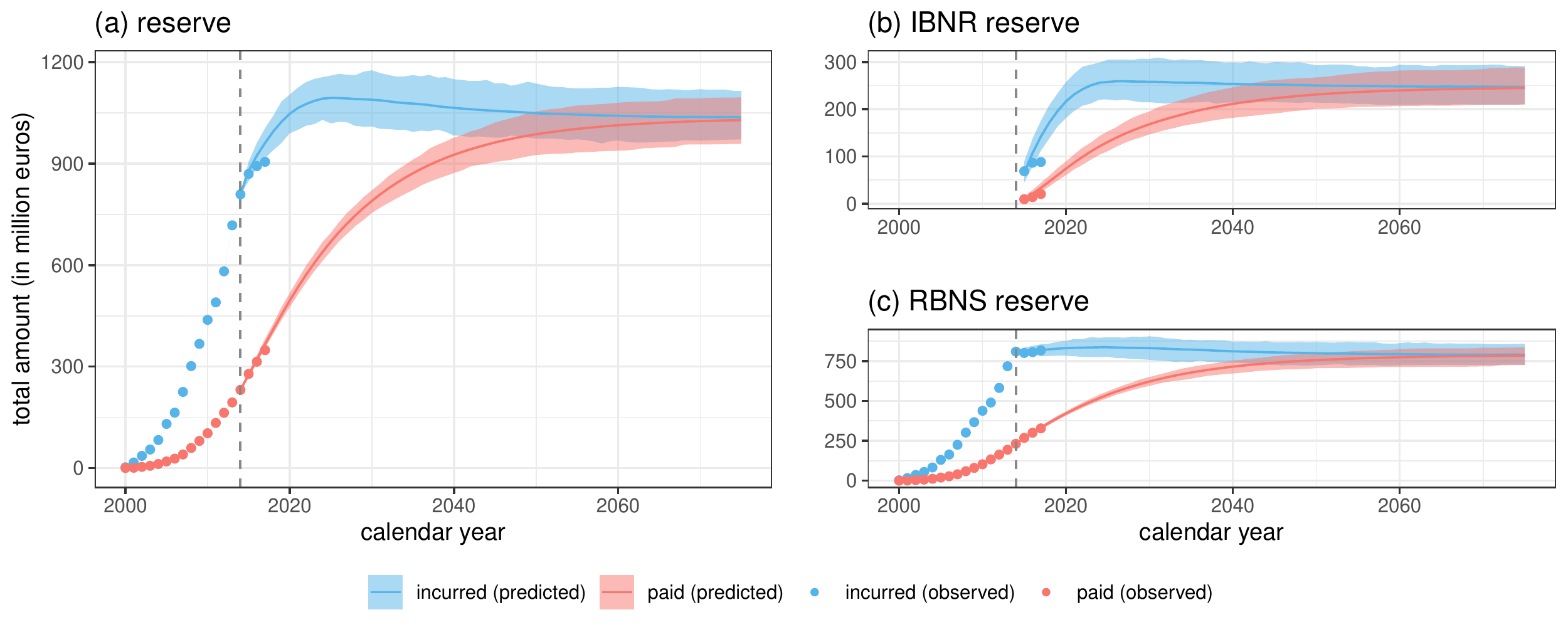}
\caption{MTPL reinsurance data set: evolution of the total amount incurred and paid for claims that occurred between January 1, 2000 and December 31, 2014 and that exceed the reporting priority of $\num{750000}$ during their development. The total reserve shown in {\bf(a)} is split into the {\bf(b)} IBNR and {\bf (c)} RBNS reserve. Simulated $95\%$ confidence intervals are shown for these amounts, with solid lines indicating expected values. Points indicate the actual out-of-time observations for calendar years 2015-2017.}
\label{figure:reserve_ground_up}
\end{figure}

Figure~\ref{figure:reserve_ground_up} shows the evolution of the total incurred and paid amounts for claims that occurred between 2000 and 2014 and exceed the reporting priority of $\num{750000}$ during their lifetime. For calendar years 2015-2017, we compare our estimates with the actual observations from the out-of-time data set. Figures~\ref{figure:reserve_ground_up_ibnr} and \ref{figure:reserve_ground_up_rbns} split the total reserve into the IBNR and RBNS reserve. For the RBNS reserve, the total amount incurred decreases slightly over time. This indicates that claim experts overestimate the expected cost of large claims when setting incurred amounts. For the total reserve (shown in Figure~\ref{figure:reserve_ground_up_total}), we estimate a sharp increase of the incurred in the first calendar years following 2014 as new claims get reported. Figure~\ref{figure:reserve_ground_up_ibnr} shows that our model overestimates the increase in the incurred, which is due to an overestimation of the number of unreported claims (not shown). In Belgium, judges use indicative tables based on mortality and discount rates to determine the compensation for bodily injury claims. In 2012, discount rates for these tables dropped from $2\%$ to $1\%$, which led claim experts to sharply increase the incurred amounts in 2013 and 2014.  This initially led to an increase in the number of reported claims, as suddenly more claims exceeded the reporting priority, followed by a decrease in reported claim counts in later years. Since adjustments for these exogenous effects can not be predicted by data driven models, expert judgement will always remain important in reserving for reinsurance. 

\begin{figure}[ht!]
\centering
\includegraphics[width = \textwidth]{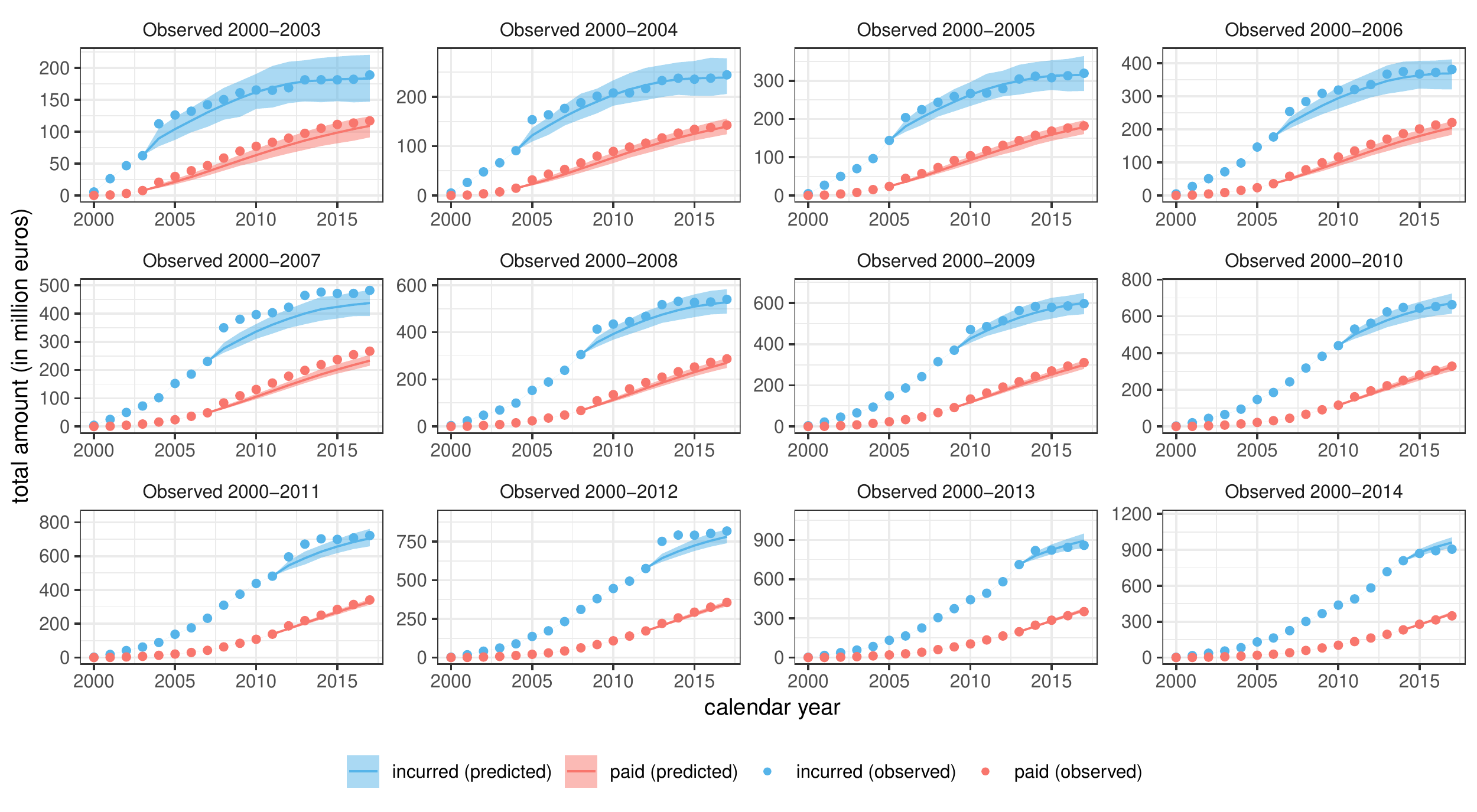}
\caption{MTPL reinsurance data set: panels show, for different observation windows, the evolution of the total amount incurred and paid until 2017 for claims that have occurred within the observation window. $95\%$ confidence intervals are shown for these amounts, with solid lines indicating expected values for years outside the observation window. Points show the actual amount incurred and paid extracted from the data until 2017.}
\label{figure:evolution_ground_up}
\end{figure}

Long delays in our reinsurance data set compel us to use most of the observed calendar years (2000-2014) for training our model, leaving only three years (2015-2017) for an out-of-time evaluation. We examine the performance of the proposed reserving model with a moving evaluation date $\tau$. We use the fitted ODM and the observed claim history at time $\tau$ to predict the future evolution of the incurred and paid amounts for claims that occurred before $\tau$. This is, however, not a true out-of-time evaluation, since we still train our ODM on the years 2000-2014. Figure~\ref{figure:evolution_ground_up} shows these evaluations of the total reserve (IBNR + RBNS) for $\tau$ ranging from 2003 to 2014. Overall, the estimated evolution of the incurred and paid amounts roughly follows the evolution recorded in our data set. The discount rate in the indicative table changed in 2002 ($4\%$ to $3\%$), 2008 ($3\%$ to $2\%$) and 2012 ($2\%$ to $1\%$). These changes cause systematic, sudden shocks in the amount incurred which can be seen on panels 2000-2007, 2000-2008 and 2000-2012 and result in an underestimation of the incurred on the short-term.

\begin{figure}[ht!]
\centering
\includegraphics[width = \textwidth]{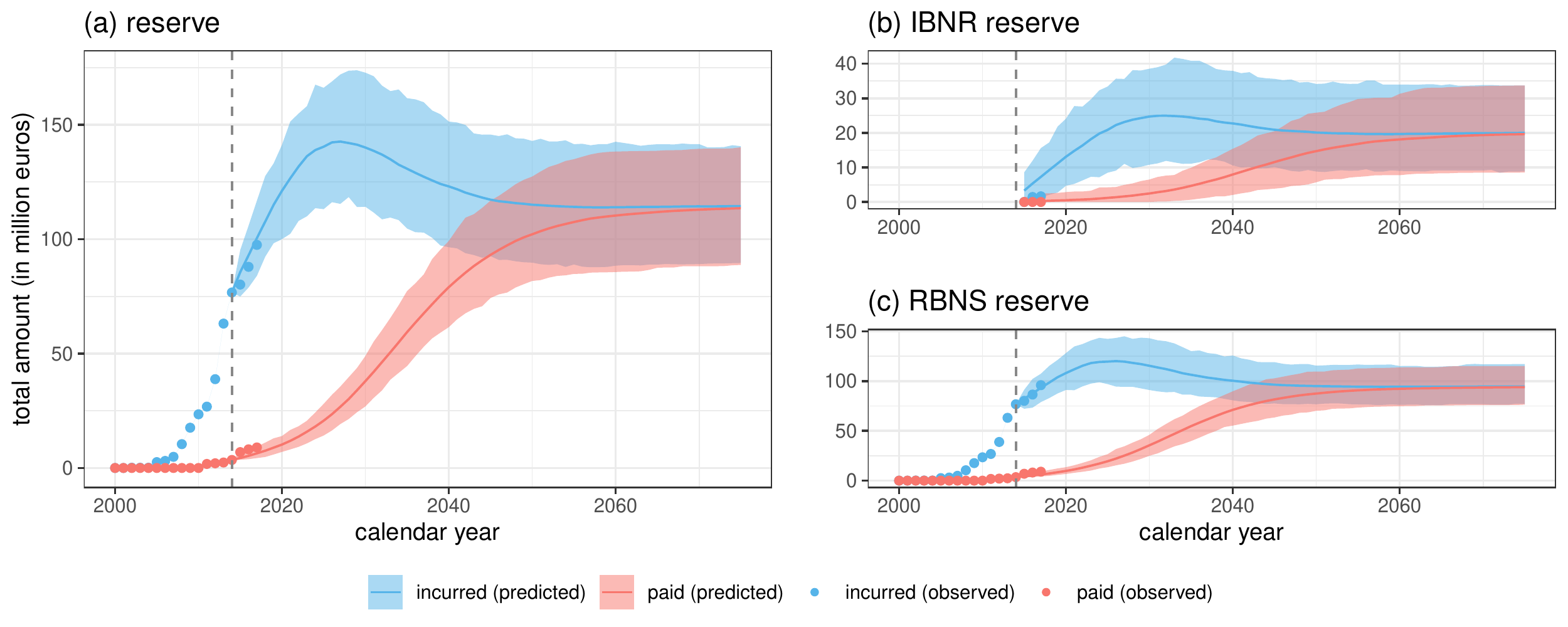}
\begin{subfigure}[t]{.9\textwidth}
\phantomcaption \label{figure:reserve_eol_total}
\phantomcaption \label{figure:reserve_eol_ibnr}
\phantomcaption \label{figure:reserve_eol_rbns}
\end{subfigure}
\caption{MTPL reinsurance data set: evolution of the aggregated amount incurred and paid between $\num{2500000}$ and $\num{5000000}$ for claims that occurred between 2000 and 2014. The {\bf(a)} total reserve is split into the {\bf(b)} IBNR and {\bf (c)} RBNS reserve. $95\%$ confidence intervals are shown for these amounts, with solid lines indicating expected values. Points indicate the actual out-of-time observations for calendar years 2015-2017.}
\label{figure:reserve_eol}
\end{figure}

\paragraph{Reserving for a portfolio of reinsurance contracts that cover the excess-of-loss}
We now focus on reserving for a portfolio of excess-of-loss contracts as considered in the pricing example of Section~\ref{section:cs_pricing}. Figure~\ref{figure:reserve_eol} shows the estimated evolution of the incurred and paid amounts within the layer of our excess-of-loss contract. Although only few payments have yet been recorded within this layer in the available reinsurance data, we can rather accurately infer the payment pattern from the general dynamics estimated with the hierarchical model of Section~\ref{section:casestudy_claim_development}. This illustrates the importance of calibrating models above a lower reporting priority in reinsurance, safeguarding a sufficient amount of data regarding the development of large claims. Where the incurred for reported claims, i.e.~the RBNS reserve, remained more or less constant when reserving from ground up (Figure~\ref{figure:reserve_ground_up_rbns}), we now observe an initial increase followed by a decrease of the total incurred within the layer of our excess-of-loss contract (Figure~\ref{figure:reserve_eol_rbns}).


\section{Conclusion} \label{section:conclusion}
We propose an occurrence and development model (ODM) for analysing the detailed claim information registered in non-life insurance portfolios. Our ODM brings valuable insights for non-life pricing as well as reserving, hereby bridging these two key actuarial tasks. We resolve the contradictions in traditional pricing literature where both actual observations as well as best estimates our used when calibrating severity models. From a reserving perspective we model the cost of IBNR claims at the level of individual policies and the future payments on RBNS claims at the level of individual claims. We illustrate our proposed methodology with two case-studies: pricing and reserving for a motor insurance as well as reinsurance portfolio. Constructing best estimates for open claims is particularly relevant and complicated in reinsurance, where reporting and settlement delays are long and claim development is uncertain. In the reinsurance setting our ODM outshines traditional methodology along two directions. First, using Jensen's inequality we demonstrate that the empirical distribution constructed from best estimates underestimates the variance of the claim severity distribution. This is then confirmed in our simulations where the claim severity distribution modelled by the ODM has a significantly larger variance than the empirical claim severity distribution based on best estimates. Second, our proposd individual reserving model captures the evolution in both paid as well as incurred amounts. Despite large uncertainties governing the development of reinsurance claims, our model is able to accurately predict the joint evolution of the paid and incurred amounts.

\section{Acknowledgements}
This work was supported by KU Leuven's research council [project COMPACT C24/15/001]; and Research Foundation Flanders (FWO) [grant number 11G4619N].

\section{Disclaimer}
This paper should not be reported as representing the views of QBE Re. The views expressed in this paper are those of the authors and do not necessarily represent those of QBE Re.

\section{Competing risks}
The authors declare none.

{
\bibliography{references}}

\appendix

\section{Layers in the hierarchical claim development models} \label{appendix:layers}

This Appendix provides a technical description of the layers used in the hierarchical development models for the MTPL insurance data set discussed in Section~\ref{section:case_insurance} and the reinsurance data set covered in Section~\ref{section:case_reinsurance}.

\subsection{Layers of the initial claim characteristics vector $I_k$}
The initial claim characteristics $\boldsymbol{I}_k$ capture the claim's information that is available when it is first reported. The set-up of $\boldsymbol{I_k}$ is tailored to each of the case-studies discussed in Section~\ref{section:casestudy}.

\subsubsection{MTPL insurance data set} 

\paragraph{1. Settlement} Indicator (yes/no) that registers whether a claim settles in the year of reporting, or not. We model this indicator with a Bernoulli distribution with logit link function. 

\paragraph{2. Payment} Indicator (yes/no) that registers whether a payment is done in the year of reporting, or not. We model this indicator with a Bernoulli distribution with logit link function. 

\paragraph{3. Increase paid}
Amount paid in the year of reporting. When there is no payment, \texttt{increase paid} is set to zero. We model \texttt{increase paid} with a gamma distribution with log link.

\paragraph{4. Initial reserve}
Reserve estimate set by the claim expert in the year of reporting. We model \texttt{initial reserve} with a gamma distribution with log link. After simulating the initial reserve, we initialize the paid and incurred amounts as follows
\begin{align*}
	\texttt{paid} &\leftarrow  \texttt{increase paid} \\
	\texttt{incurred} &\leftarrow \texttt{initial reserve} + \texttt{paid}.
\end{align*}

\subsubsection{MTPL reinsurance data set} 

\paragraph{1. Excess incurred} This layer registers the difference between the claim's incurred amount as set by the insurer and the reporting priority in the year that the claim is reported. This excess incurred is positive since claims are only reported by the insurer when their incurred amount exceeds the reporting priority. Furthermore the excess incurred will be at least 100 euro as a result of data preprocessing. When modelling the excess incurred we first apply a power transform and then model the transformed outcome with a truncated Gaussian distribution, i.e.
$$ \texttt{excess incurred}^{p} \sim \text{trunc.Gaussian}(\mu, \sigma, T = 100^{p}).$$
For the data set analyzed in Section~\ref{section:case_reinsurance} the power $p$ was calibrated as $0.117$. After simulating \texttt{excess incurred} we compute the \texttt{incurred} as
$$\texttt{incurred} = \texttt{reporting priority} + \texttt{excess incurred}.$$

\paragraph{2. Payment} Indicator (yes/no) registering whether the insurer made any claim payments before or in the year of reporting the claim to the reinsurer. We model this indicator with a Bernoulli distribution with logit link function.

\paragraph{3. Pct paid} When there is a payment, we model the amount paid at reporting as a percentage of the incurred at reporting. After simulating this layer the paid amount and the reserve are computed as
\begin{align*}
	\texttt{paid} &=  \begin{cases}
	0 & \texttt{payment} = \text{no} \\
	\texttt{incurred} \cdot \texttt{pct paid} & \texttt{payment} = \text{yes}
	\end{cases}, \\
	\texttt{reserve} &= \texttt{incurred} - \texttt{paid}.
\end{align*}

Modelling the percentage instead of the paid amount has the advantage that the condition $\texttt{paid} \leq \texttt{incurred}$ is automatically satisfied. We model \texttt{pct paid} by first applying a logit transform and then assuming a Gaussian distribution for the transformed variable, i.e.~
$$\text{logit}(\texttt{pct paid}) \sim \texttt{Gaussian}(\mu, \sigma).$$

\subsection{Layers of the update vectors $U_k^{j}$}

The set-up of $\boldsymbol{U}_k^{j}$ is almost identical for both case-studies in Section~\ref{section:casestudy}.

\paragraph{1. Settlement} Indicator (yes/no) that registers whether a claim settles in the current development year, or not. We model this indicator with a Bernoulli distribution with logit link function.

\paragraph{2. Payment} Indicator (yes/no) that registers whether the insurer made a payment in the current development year, or not. In the reinsurance case study, this indicator is \texttt{yes} when the payment size exceeds 100 euro. We model this indicator with a Bernoulli distribution with logit link function.

\paragraph{3. Increase paid} Amount paid in the current development year. When there is no payment, \texttt{increase paid} is set to zero. If there is a payment, we first apply a power transform on this variable in the reinsurance case study, and then assume a truncated Gaussian distribution for the transformed variable. In the insurance case study we use a gamma distribution because claim sizes are smaller. After simulating \texttt{increase paid}, we increase the amount paid by the size of the new payment and subtract this payment from the outstanding reserve, i.e.
\begin{align*}
	\texttt{paid} &\leftarrow \texttt{paid} + \texttt{increase paid}, \\
	\texttt{reserve} &\leftarrow \text{min}(0, \texttt{reserve} - \texttt{increase paid}),\\
	\texttt{incurred} &\leftarrow \texttt{paid} + \texttt{reserve}.
\end{align*}

\paragraph{4. Change reserve} Indicator (yes/no) registering whether the reserve changes in the current development year. We only model reserve changes when the claim does not settle in the current year. In the year of settlement the reserve is deterministically set to zero. In the reinsurance case-study this indicator is only triggered by changes of at least 100 euro as a result of the data preprocessing step. This layer is modelled with a Bernoulli distribution with logit link function.

\paragraph{5. Reserve is zero} Indicator (yes/no) registering whether the reserve drops to zero. This layer is modelled conditionally on $\texttt{change reserve} = \text{yes}$ and $\texttt{reserve} \neq 0$. This layer is modelled with a Bernoulli distribution with logit link function. This layer is not included in the insurance case study, since there the reserve is always larger than zero when the claim has not yet settled. 

\paragraph{6. Change reserve pos} Indicator (yes/no) registering whether the reserve increases in the current year. This layer is modelled conditionally on $\texttt{change reserve} = \text{yes}$, $\texttt{reserve is zero} = \text{no}$ and $\texttt{reserve} \neq 0$. When $\texttt{change reserve} = \text{yes}$ and $\texttt{reserve} = 0$, this layer is always set to yes as the reserve cannot decrease.

\paragraph{7. Increase reserve}
Nominal increase in the reserve conditional on an increase in the reserve. In the reinsurance case study, these increases are modelled with a truncated Gaussian distribution after applying a power transformation. A gamma GLM with log link is used in the insurance case study.

\paragraph{8. Pct decrease reserve}
Percentage decrease in the reserve conditional on a decrease in the reserve. As a result of pre-processing, this percentage is lower bounded by $\frac{100}{\texttt{reserve}}$ in the reinsurance case study. In modelling, we first apply a logit transform to this percentage and then model the transformed outcome with a truncated Gaussian distribution. After simulating this layer, we update the reserve and incurred as
\begin{align*}
	\texttt{reserve} &\leftarrow \begin{cases}
	0 & \texttt{settlement} = \text{yes}, \\
	\texttt{reserve} & \texttt{change reserve} = \text{no}, \\
	0 & \texttt{reserve is zero} = \text{yes}, \\
	\texttt{reserve} + \texttt{increase reserve} & \texttt{change reserve pos} = \text{yes}, \\
	\texttt{reserve} \cdot (1-\texttt{pct decrease reserve}) & \text{otherwise}.
	\end{cases} \\
	\texttt{incurred} &\leftarrow \texttt{paid} + \texttt{reserve}.
\end{align*}

\section{Simulation strategies for claim severities}\label{sec:sim_strat_sev}

We illustrate both simulation strategies discussed in Section~\ref{section:pricing} on the reinsurance data set. Our goal is to simulate paths from the severity distribution of a claim resulting from portfolio A, with occurrence year $2015$.

\begin{figure}[ht!]
\centering
\includegraphics[width = \textwidth]{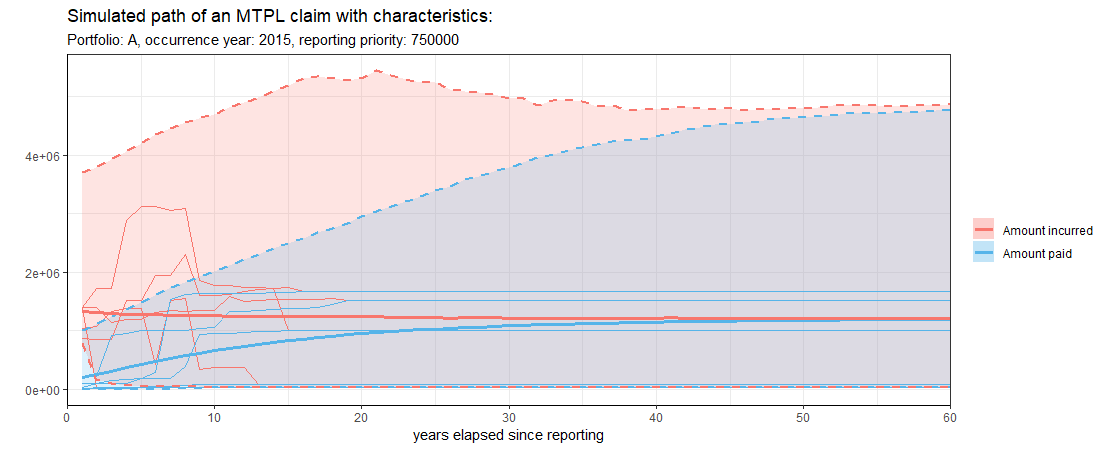}
\caption{MTPL reinsurance data set: simulated evolution of the amount incurred and amount paid for a new claim from portfolio \texttt{A} that occurs in year 2015 and is reported at a priority of $\num{750000}$. Thick solid lines show the average amount paid and incurred, while dashed lines indicate the $95\%$ confidence intervals for these amounts. Thin lines visualize the evolution of 5 randomly selected scenarios.}
\label{figure:simulated_evolution_single_claim}
\end{figure}

\paragraph{Simulating paths for a new claim}
We simulate $\num{20000}$ paths for the development of a new claim from policy $\texttt{A}$ that occurred in $2015$. Hereto, we follow the steps outlined in Algorithm~\ref{algorithm:full_simulation}. The occurrence year 2015 is not used in our claim development model, but is required for deflating the simulated paths. Figure~\ref{figure:simulated_evolution_single_claim} visualizes the evolution of the paid and incurred amounts as obtained over these $\num{20000}$ paths. Solid lines indicate the average paid and incurred amounts, whereas the dashed lines bound the $95\%$ confidence interval. At reporting the incurred exceeds $\num{750000}$ for all simulated paths. However, soon after reporting the lower bound for the incurred drops to zero as some of these paths will settle without payment, representing the case where the claim is not eligible for compensation within the portfolio. This is a common scenario for large motor insurance claims, where often many parties and hence insurers are involved in an accident and it may initially not be clear which insurer should reimburse the claim. After 15 years have elapsed, i.e.~the observation window of our training data set, many simulated claims are still open. This is visible in Figure~\ref{figure:simulated_evolution_single_claim} by the large difference between the paid and incurred amounts after 15 years. Supported by the low importance of the covariate \texttt{number of elapsed years since reporting} in all layers of the hierarchical development model (Figure~\ref{figure:variable_importance_gbm}), we extrapolate our model and simulate the development up to 60 years after the reporting of the claim. After 60 years almost all paths have settled and the paid amount has converged towards the amount incurred. A settlement delay of 60 years may seem long, but occurs in practice when victims are compensated via lifelong periodic payments. The empirical distribution of the total amount paid after 60 years is our simulated severity distribution for a new claim from policy $\texttt{A}$ that occurs in 2015 and is reported with a priority of $\num{750000}$. 

\begin{figure}[ht!]
\centering
\includegraphics[width = \textwidth]{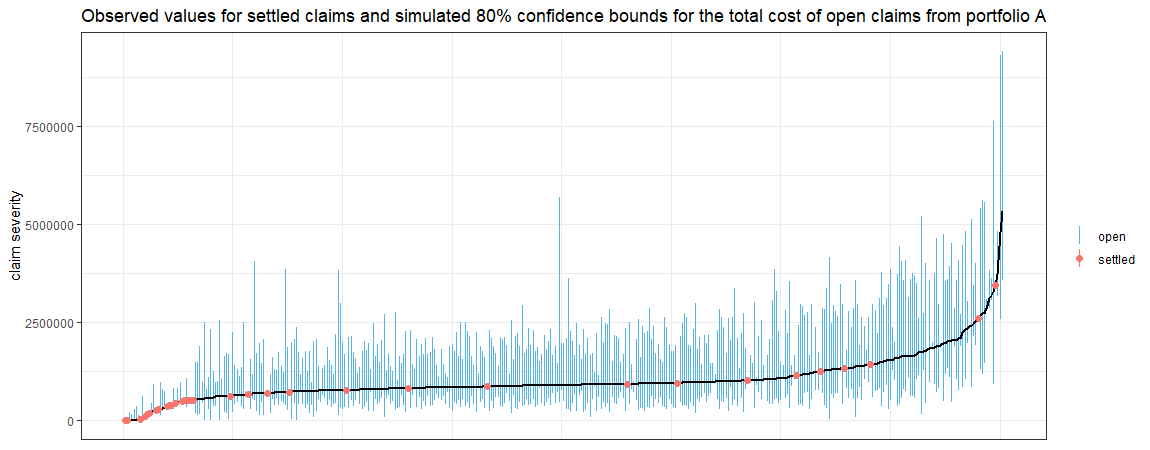}
\caption{MTPL reinsurance data set: $80\%$ confidence intervals for 401 observed claims from portfolio $\texttt{A}$ based on 200 simulations per open claim. The actual reporting priority of portfolio $\texttt{A}$ is less than $\num{750,000}$ and as a result the incurred of some simulated paths never exceed the chosen reporting priority of $\num{750000}$. These paths are omitted from the results. Observed claims are sorted by median loss, which is indicated with a solid black line.}
\label{figure:simulated_ci_claim}
\end{figure}

\paragraph{Simulating future paths for open claims}
Alternatively, we focus on the observed claim data from insurance portfolio $\texttt{A}$. By the end of 2014, we observe 401 claims from this portfolio of which 33 are closed and 368 are open. We simulate 200 future development paths for each open claim. Figure~\ref{figure:simulated_ci_claim} shows the total amount paid for settled claims and a $80\%$ confidence interval for the amount paid at settlement based on the 200 simulated paths per open claim. Compared to Figure~\ref{figure:simulated_evolution_single_claim}, we show $80\%$ instead of $95\%$ confidence intervals, since the scale on the vertical axis is heavily impacted by extreme outcomes registered for individual claims. Claims are sorted by median severity, indicated with a solid black line. The distribution of the simulated ultimate claim amount is heavily right skewed with the median near the lower end of the confidence interval. By maximizing the likelihood in~\eqref{eq:likelihood_weighted_paths} a severity distribution can be estimated from the observed ultimate claim sizes of settled claims and the simulations for the ultimate claim amounts of open claims. In the reinsurance case-study, we use the empirical cumulative distribution function (ecdf) as a non-parametric estimator for the claim severity distribution. In the construction of this ecdf the observed outcomes (settled claims) get a weight of 1 and simulated paths (open claims) each get a weight of $\frac{1}{\texttt{number of simulations}}$. 

\end{document}